\documentclass[%
 aip,
 sd,%
 amsmath,amssymb,
 reprint,%
]{revtex4-2}

\usepackage{multirow}
\usepackage{graphicx}
\usepackage{bm}
\usepackage{hyperref}

\usepackage{graphicx, subfigure, textcomp, amssymb, mathrsfs}

\usepackage[version=3]{mhchem} 
\usepackage{chemarr}

 \makeatletter
\newcommand\xleftrightarrow[2][]{%
  \ext@arrow 9999{\longleftrightarrowfill@}{#1}{#2}}
\newcommand\longleftrightarrowfill@{%
  \arrowfill@\leftarrow\relbar\rightarrow}
\makeatother

\usepackage{tikz}
\usepackage{tikz-cd}
\usetikzlibrary{matrix, calc, arrows}
\usepackage{circuitikz,tikz}
\usepackage{xr}
\usepackage{epstopdf}

\begin{document}

\title{Non-Debye impedance and relaxation models for dissipative electrochemical capacitors}

\author{Anis Allagui$^*$}
\email{aallagui@sharjah.ac.ae}
\affiliation{Dept. of Sustainable and Renewable Energy Engineering, University of Sharjah, Sharjah, P.O. Box 27272, United Arab Emirates}
\altaffiliation[Also at ]{Center for Advanced Materials Research, Research Institute of Sciences and Engineering, University of Sharjah, Sharjah,, P.O. Box 27272,  United Arab Emirates}
\affiliation{Dept. of Mechanical and Materials Engineering, Florida International University, Miami, FL33174, United States}

\author{Hachemi Benaoum} 
\affiliation{
Dept. of Applied Physics and Astronomy, 
University of Sharjah, PO Box 27272, Sharjah, United Arab Emirates 
}

\author{Ahmed S. Elwakil}
\affiliation{Dept. of Electrical and Computer Engineering, University of Sharjah, Sharjah, P.O. Box 27272, United Arab Emirates}
\affiliation{Nanoelectronics Integrated Systems Center, Nile University, Cairo 12588, Egypt}
\affiliation{Dept. of Electrical and Computer Engineering, University of Calgary, Calgary, Alberta T2N 1N4, Canada}

\author{Mohammad Alshabi}
\affiliation{Dept. of Mechanical and Nuclear Engineering, University of Sharjah, PO Box 27272, Sharjah, United Arab Emirates}

\begin{abstract}

Electrochemical capacitors are a class of energy devices in which complex mechanisms of accumulation and dissipation of electric energy take place when connected to a charging or discharging power system. Reliably  modeling  their frequency-domain and time-domain behaviors is very important for their proper design and integration in engineering applications, knowing that  electrochemical capacitors in general exhibit anomalous tendency that cannot be adequately captured with traditional integer-order-based models. In this study we first review some of the widely used fractional-oder models for the description of impedance and relaxation functions  of dissipative resistive-capacitive system, namely the Cole-Cole, Davidson-Cole, and Havriliak–Negami models. 
 We then propose and derive new $q$-deformed  models based on  modified evolution equations for the charge or voltage when the device is discharged into a parallel resistive load. We verify our results on  anomalous spectral impedance response and time-domain relaxation   data for voltage and  charge obtained from a commercial supercapacitor. 
 


\end{abstract}

\maketitle

\section{Introduction}

Electrochemical capacitors are energy storage devices relying on the very large electric double-layer capacitance at the porous electrode/electrolyte interface for electrostatic charge storage, and/or  fast and reversible faradic redox reactions for pseudocapacitive storage \cite{zhao2021electrochemical, zhao2021structural}. 
Their  capacitance, potential window, energy and power capabilities  are dependent on several geometrical and physical parameters, including the surface area, type and microstructural complexity of the electrodes being used, interfacial charge absorption/transfer, ions electrodiffusion and migration dynamics, types of supporting electrolyte and ionic strength, etc \cite{yang2022understanding}. 
The direct identification of their physical parameters and characterization of underlying microscopic processes occurring in such systems is quite challenging. It often requires sophisticated systems of instrumentation with high in situ accuracy and resolution levels. However, this can be circumvented to a certain extent by analyzing instead the  measurements of macroscopic quantities (current and voltage) in time-domain relaxation experiments or from frequency-domain impedance or admittance \cite{fracorderreview}.  
Such information allows one to gain useful insight into the microscopic processes taking place in the system without going into the unnecessary details. 
 Furthermore, it is important for practical purposes to be able to describe a system in the frequency domain when time-domain data are available, and vice versa, describing the system in the time domain when the data available are in the frequency domain using the appropriate transformations \cite{ieeeted, allagui2021inverse, acs2}. Of  course, it is understood that frequency-domain formulation becomes inappropriate when the system under consideration involves nonlinear, local-in-time effects. This makes time-domain formulation  more convenient in this context. 
 
The simplest type of electrochemical capacitors, i.e. electric double-layer capacitors (EDLCs), can be viewed as an ideally non-polarizable two-electrode cell system, where faradaic reactions occur very fast and the charge transfer resistance is negligible. They can be modeled using a parallel $RC$ circuit representing the bulk resistance and bulk capacitance of the device, which is usually a good starting point to model their behavior \cite{yang2022understanding}. We assume  the series resistance to be negligible for modeling convenience.  The    relaxation response of an $RC$ circuit is a decaying exponential function with time (with a single characteristic relaxation constant $\tau$), which is known to be the eigenfunction of the  first-order time derivative operator, i.e. the   evolution  equation $ {\mathrm{d}\rho(t)}/{\mathrm{d}t}+\tau^{-1}{\rho(t)}=0$ (Eq.\;\ref{eqrho} below).   
However, it is becoming evident from many experimental data that the relaxations of porous electrodes and complex electrochemical capacitors in general are rather nonexponential, power-law-like profiles \cite{10.1149/1945-7111/ac621e}. From frequency-domain measurements, the corresponding impedance does not show the expected semi-circle of imaginary vs. real parts of an ideal $RC$ circuit \cite{eis}.  These observations indicate that from a macroscopic point of view the system cannot be viewed simply as the collection of many ideal, non-interacting subsystems. 
 
 Anomalies in the electrical response and frequency dispersion of materials and devices in general are characteristic features of disorder, spatial heterogeneity (e.g. fractal and porous structures), and wide spectrum of relaxation times. The treatment of such type of data usually requires extending the traditional kinetic equation $ {\mathrm{d}\rho(t)}/{\mathrm{d}t}+\tau^{-1}{\rho(t)}=0$ using tools borrowed from fractional calculus. This leads to the well-known   fractional impedance models of  Cole-Cole \cite{cole1941dispersion}, Davidson-Cole \cite{davidson1951dielectric} and Havriliak–Negami \cite{havriliak1966complex}  which can viewed as extension of the Debye model by introducing one or two fractional exponents\;\cite{hilfer2002analytical, goychuk2007anomalous, cpe}. Their corresponding time-domain relaxation dynamics are expressed in terms of the Mittag-Leffler and the Fox's $H$-functions\;\cite{hilfer2002analytical}. 
 In this study, after reviewing  the above mentioned fractional models,  we propose and derive new deformed dynamic models for resistive-capacitive systems using the linear and nonlinear time $q$-derivatives introduced in the kinetic equation. 
The difficulty that arises when attempting to assign a physical interpretation related to non-local (global) fractional time derivatives (i.e. hereditariness that takes into account the past activities of the system up to the current time, Fig.\;\ref{fig1}) associated with the   Cole-Cole, Davidson-Cole  and Havriliak–Negami   is not an issue with the $q$-derivatives. 
The solutions for the time-domain relaxation are provided in terms of the $q$-exponential and logistic functions. The associated frequency-domain responses are also derived and discussed.  
 
\begin{figure}[t]
\begin{center}
\includegraphics[width=.45\textwidth]{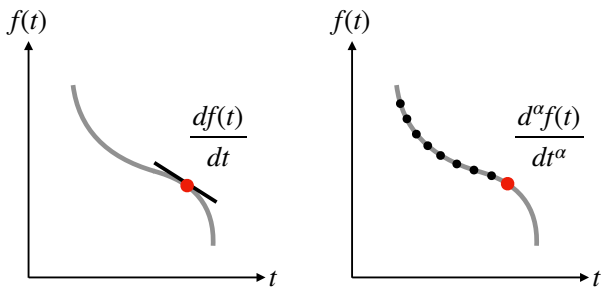}
\caption{Graphical comparison between first-order, local derivative of a function $f(t)$ vs. its fractional-order, non-local derivative that takes into account not just the immediate past of the function proceeding the instant $t$, but the whole past of the function.}
\label{fig1}
\end{center}
\end{figure}

The remainder of the paper is organized as follows. We first present in Section\;\ref{RC} a short overview of the   time- and frequency-domain Debye models for the case of an $RC$ system. 
In Section\;\ref{FO}, we outline the different forms of fractional kinetic models resulting in nonexponential relaxation functions as well as their associated impedance functions of the Cole-Cole, Davidson-Cole and Havriliak–Negami type. Subsequently, the $q$-deformed models are derived and discussed in Section\;\ref{q}.   
Frequency-domain and  time-domain experimental results measured  on a commercial supercapacitor are analyzed and modeled in Section\;\ref{RD}.
 

\section{Classical $RC$ model\label{RC}}

First, we recall that the impedance of a parallel $RC$ circuit model is given by the ratio of the Laplace transforms (defined as $\mathcal{L} (f(t);s) = F(s) =\int_0^{\infty} f(t)e^{-st} \mathrm{d}t$) of time-domain voltage $v(t)$ and current $i(t)$ as:
\begin{equation}
Z(s) = \frac{\mathcal{L}{(v(t))}}{\mathcal{L}(i(t))} =\frac{V(s)}{I(s)} = \frac{R}{1+s \tau}
\label{eq1}
\end{equation}
 where $s=j\omega $ and $\tau=R C$ is a characteristic relaxation time.
Up to a constant, Eq.\;\ref{eq1} has the same form as the normalized complex susceptibility model of dielectrics provided by Debye,  
i.e.
$\chi (s) = ({1+s \tau_D})^{-1}$.  
Its corresponding response function obtained by inverse Laplace transform  of the normalized complex susceptibility  is 
$\phi(t) = \mathcal{L}^{-1} (\chi(s);t) = \tau_D^{-1}   e^{-t/\tau_D}$ ($t\geqslant 0$).  
 The relaxation function defined as $\Psi(t)= 1 -\mathcal{L}^{-1}(\chi(s)/s; t)$ is   equal to $ e^{-t/\tau_D}$ \cite{garrappa2016models}. 
The half lifetime characteristic representing the time at which  $\Psi(t)$ reaches half of its initial value is  $ \tau_D  \ln(2) \approx 0.693 \, \tau_D$. 

In the same way, the time-domain  voltage decay of the capacitor, that we denote $\rho(t)$, corresponding to its discharge into a parallel resistance from an initial charge $\rho(0)=1$ is given by:
 \begin{equation}
\textcolor{black}{\rho(t) = e^{-t/\tau}}
\label{eq2}
\end{equation}  
Eq.\;\ref{eq2} represents also the evolution of the electrical charge  as the pre-charged capacitor self-discharges into the parallel resistance. 
It is also the solution of the linear  integer-order differential equation:
\begin{equation}
\frac{\mathrm{d}\rho(t)}{\mathrm{d}t}+ \tau^{-1}{{\rho(t)}}=0
\label{eqrho}
\end{equation} 
This is the equation of the standard kinetic model in which the rate of change $\mathrm{d}\rho(t)/\mathrm{d}t$ of a certain physical quantity $\rho(t)$ as it is approaching equilibrium by the action of an external excitation is proportional to the  quantity itself.  
Generalizations of Eq.\;\ref{eqrho} using different forms of fractional-order, instead of integer-order, time derivatives of $\rho(t)$   are presented in Section\;\ref{FO}. Models  obtained when elevating   $\rho(t)$ to a power of $q\in \mathbb{R}$ while maintaining the integer-order derivative in Eq.\;\ref{eqrho} are presented and discussed in Section\;\ref{q}

\section{Fractional-order models\label{FO}}

 Generalization of the  Debye models   for complex susceptibility and relaxation function in dielectrics include the nonexponential Cole-Cole \cite{cole1941dispersion}, Davidson-Cole \cite{davidson1951dielectric} and Havriliak–Negami \cite{havriliak1966complex} models.
  Equivalently, the normalized Cole-Cole impedance model  is given by \cite{cole1941dispersion}: 
\begin{equation}
\textcolor{black}{Z_{\alpha} (s) = \frac{1}{1 + (s\tau_{\alpha}) ^{\alpha}}} \quad (0<\alpha \leqslant 1)
\label{eqCC}
\end{equation}
 which admits the  time-domain voltage or charge relaxation   in terms of the Mittag-Leffler (ML) function as \cite{garrappa2016models,10.1149/1945-7111/ac621e}:
\begin{align}
\rho_{\alpha}(t) &= 1-(t/\tau_{\alpha})^{\alpha} E^1_{\alpha,\alpha+1}  \left[ -(t/\tau_{\alpha})^{\alpha} \right] \\
&= E_{\alpha,1}^1\left[ -(t/\tau_{\alpha})^{\alpha} \right]
\label{eq41}
\end{align}
where the   initial condition is $\rho_{\alpha}(0)=1$.  
This result is obtained from Prabhakar integral \cite{prabhakar1971singular, saxena2004generalized}:
\begin{equation}
  \int \limits_0^{\infty}  t^{\beta-1}  {E}_{\alpha,\beta}^{\gamma} \left( -at^{\alpha}\right) e^{-st} dt =  \frac{s^{-\beta}}{(1+as^{-\alpha})^{\gamma}} = \frac{s^{\alpha\gamma-\beta}}{(s^{\alpha}+a)^{\gamma}} 
  \label{e5}
\end{equation}
where 
\begin{equation}
{E}_{\alpha,\beta}^{\gamma} ( z ) := \sum\limits_{k=0}^{\infty} \frac{(\gamma)_k}{\Gamma(\alpha k + \beta)} \frac{z^k}{k!} \quad (\alpha,\beta, \gamma \in \mathbb{C}, \mathrm{Re}({\alpha})>0)
\label{eqML}
\end{equation}
 (with $(\gamma)_k = \gamma(\gamma+1)\ldots(\gamma+k-1) =\Gamma(\gamma+k)/\Gamma(\gamma)$ being the Pochhammer symbol and $\Gamma(z)$ is the gamma function) is the three-parameter ML function. At the limit of $\alpha \to 1$, one readily recovers the Debye function ($E_1(-z)=e^{-z}$). The half lifetime characteristic can be obtained from the inverse of the ML function \cite{zeng2015global, hilfer2006computation}. 
 We note that  Eq.\;\ref{eq41} can be expressed in terms of Fox's $H$-function (see Appendix\;\ref{H}) as \cite{hilfer2002analytical}:
 \begin{equation}
\rho_{\alpha}(t) = H^{1,1}_{1,2} \left[(t/\tau_{\alpha})^{\alpha}|^{(0,1)}_{(0,1),(0,\alpha)} \right]
\end{equation} 
 Eq.\;\ref{eq41} is also the solution to \cite{khamzin2014justification}:
\begin{equation}
\frac{\mathrm{d}\rho_{\alpha}(t)}{\mathrm{d}t} +\tau_{\alpha}^{-\alpha} \, _0D_t^{1-\alpha}\rho_{\alpha}(t)  =0
\label{eq217}
\end{equation}
 with $\rho_{\alpha}(0)=1$, and  
 $_0D_t^{1-\alpha} f(t) = (\mathrm{d}/\mathrm{d}t)_0D_t^{-\alpha} f(t)$ denotes  the fractional derivate in the Riemann-Liouville sense  with:
  \begin{equation}
 _0D_t^{-\alpha} f(t) =  \frac{1}{\Gamma(\alpha)} \int_0^t    {(t-\tau)^{\alpha-1}}   f(\tau)   d\tau
 \label{eqRL1}
\end{equation}
being   the Riemann-Liouville fractional integral. 
 Eq.\;\ref{eq41} is also the solution to  \cite{garrappa2016models}:
\begin{equation}
 {{}_0^CD_t^{\alpha}  \rho_{\alpha}(t) + \tau_{\alpha}^{-1}{\rho_{\alpha}(t) } }=0
 \label{eqFKCC}
\end{equation}
 where  ${}^C_0 D_t^{\alpha} f(t)$ is here defined here in the  Caputo sense  by \cite{podlubny1998fractional}:      
\begin{equation}
 {}^C_0D_t^{\alpha} f(t) =  \frac{1}{\Gamma(1-\alpha)} \int_0^t  {(t-\tau)^{-\alpha}} \frac{\mathrm{d}f(\tau)}{\mathrm{d}\tau}       d\tau
\label{eqRL}
\end{equation} 
Eq.\;\ref{eq217} or Eq.\;\ref{eqFKCC} are one way of generalizing the integer-order rate equation given by Eq.\;\ref{eqrho} by having global, non-local time derivative of $\rho_{\alpha}(t)$  proportional to the quantity itself, and thus including memory effects \cite{khamzin2014justification, nigmatullin1984theory, memoryAPL, memQ, allagui2021possibility}.   

On the other hand, the (normalized) Davidson-Cole impedance model is given by \cite{davidson1951dielectric}:
\begin{equation}
\textcolor{black}{Z_{\beta}(s) = \frac{1}{(1+ s \tau_{\beta})^{\beta}}} \quad  (0<\beta \leqslant 1)
\label{eqDC}
\end{equation}
and its corresponding relaxation function (with the use of Eq.\;\ref{e5}) is \cite{hilfer2002analytical, garrappa2016models}:
\begin{equation}
\textcolor{black}{\rho_{\beta}(t) =1- (t/\tau_{\beta})^{\beta} E_{1,\beta+1}^{\beta} (-t/\tau_{\beta})}
\label{eq12}
\end{equation}
which can also be expressed as \cite{hilfer2002analytical}:
\begin{equation}
\rho_{\beta}(t)=  \frac{\gamma(\beta,t/\tau_{\beta})}{\Gamma(\beta)} = 1- \frac{1}{\Gamma(\beta)}  H^{1,1}_{1,2} \left[(t/\tau_{\beta})|^{(1,1)}_{(\beta,1),(0,1)} \right]
\end{equation} 
where 
$\gamma(a,z) = \int_z^{\infty} x^{a-1} e^{-x} dx$  
is  the complementary incomplete gamma function.  
The corresponding differential equation for $\rho_{\beta}(t)$ with the usual initial condition $\rho_{\beta}(0)=1$  is in this case \cite{khamzin2014justification, rosa2015relaxation}:
\begin{equation*}
\frac{\mathrm{d}\rho_{\beta}(t)}{\mathrm{d}t} +
\tau_{\beta}^{-\beta}
\frac{\mathrm{d}}{\mathrm{d}t}
\left\{ e^{-t/\tau_{\beta}} \mathrm{\mathbf{E}}^1_{\beta,\beta,\tau_{\beta}^{-\beta},0+} e^{\tau/\tau_{\beta}} \rho_{\beta}(t) \right\}
 = 0
\end{equation*} 
with the use of the Kilbas-Saigo-Saxena integral operator \cite{kilbas2004generalized}:
\begin{align}
\mathrm{\mathbf{E}}_{\rho,\mu,\omega,a+}^{\gamma}  f(t) 
&= \int_a^t (t-\tau)^{\mu-1} E_{\rho,\mu}^{\gamma} [\omega(t-\tau)^{\rho}] f(\tau) d\tau \\
&= \sum\limits_{m=0}^{\infty} \frac{\omega^m (\gamma)_m }{m!}  \,_0D_t^{-\rho m + \mu} f(t)
\end{align}

Finally, 
the Havriliak-Negami impedance function is a further generalizations of the Cole-Cole and Davidson-Cole  models, and is given by \cite{havriliak1966complex}:
\begin{equation}
\textcolor{black}{Z_{H}(s) = \frac{1}{\left(1+ (s \tau_H)^{\alpha}\right)^{\beta}}} \quad ( 0< \alpha, \beta  \leqslant 1)
\label{eqH}
\end{equation}
For  $\beta=1$ we recover the Cole-Cole model, while for $\alpha=1$ the  Davidson-Cole model is recovered, and with $\alpha=\beta=1$ we end up with the Debye model. 
 From Eq.\;\ref{e5}, the corresponding relaxation function for the Havriliak-Negami model can be expressed in terms of the Mittag-Leffler function 
  as \cite{garrappa2016models}:
 \begin{equation}
\textcolor{black}{\rho_{H}(t)= 1- (t/\tau_H)^{\alpha \beta} E_{\alpha,\alpha\beta+1}^{\beta}\left[- (t/\tau_H)^{\alpha} \right]
}
\label{eqHH}
\end{equation}
   or in terms of the $H$-function  as \cite{hilfer2002analytical,weron2005havriliak, garrappa2016models}:
 \begin{equation}
{\rho_{H}(t) = 1 - \frac{1}{\Gamma(\beta)} H_{1,2}^{1,1} \left[ {(t/\tau_H)}^{\alpha} |^{(1,1)}_{(\beta,1).(0,\alpha)} \right]}
\end{equation}
The differential equation for $\rho_{H}(t)$ is \cite{khamzin2014justification, rosa2015relaxation}: 
\begin{equation}
\frac{\mathrm{d}\rho_{H}(t)}{\mathrm{d}t}
 +
\sum\limits_{k=0}^{\infty} \tau_H^{-\alpha\beta(k+1)} \frac{\mathrm{d}}{\mathrm{d}t}  \mathrm{\mathbf{E}}^{\beta(k+1)}_{\alpha,\alpha\beta(k+1),-\tau_H^{-\alpha},0+} \rho_{H}(t)
 = 0
 \label{eqEEqHN}
\end{equation}
  
 Note that apart from the above time-domain relaxation functions, their frequency-domain counterparts can also be expressed in terms of the $H$-function (see   Hilfer \cite{hilfer2002h}).

\section{$q$-deformed models\label{q}}

Now if the first-order time derivative of $\rho(t)$ in Eq.\;\ref{eqrho} is taken instead proportional to a power of   $\rho(t)$  such that:
\begin{equation}
 \frac{\mathrm{d \rho}_{q_1}(t)}{\mathrm{d}t} +{ \left[\frac{\rho_{q_1}(t)}{\tau_{q_1}}\right]^{q_1}} =0
\label{eq3}
\end{equation}
where $q_1$ is a real parameter,  then the solution for $\rho_{q_1}(t)$ with $\rho_{q_1}(0)=1$  admits the power-law behavior:
\begin{equation}
\textcolor{black}{\rho_{q_1}(t)=  [1 - (1-{q_1}) (t/\tau_{q_1})]^{\frac{1}{1-{q_1}}} =   e_{q_1}^{- t/\tau_{q_1}} }
\label{eq4}
\end{equation}
where $e_q^x$ is defined as the $q$-exponential function \cite{borges1998q}. 
We note here some of its  properties:  
(i)   for  $q<1$, $e_q^x=0$ for $x<-1/(1-q)$ and $e_q^x=\left[1+(1-q)x \right]^{1/(1-q)}$ for $x\geqslant -1/(1-q)$, 
  (ii)  for $q=1$, $e_q^x=e^x$ for $\forall y$, and 
  (iii) for $q>1$, $e_q^x= \left[1+(1-q)x \right]^{1/(1-q)}$ for $x < 1/(q-1)$ \cite{borges1998q}.
 Its Taylor series expansion is:
  \begin{equation}
 {e_q^{-x}} = e^{-x} \left[ 1+ \frac{1}{2}(q-1)x^2 - \frac{1}{3}(q-1)^2 x^3 + \ldots \right]
\label{eqT}
\end{equation} 
from which it is clear that at the limit $q\to 1$, one recover the ordinary exponential law $\rho(t)= e^{- t}$ (i.e. Eq.\;\ref{eq2}).  

  We note also that the $q$-exponential and   $q$-Gaussian distributions  are the functions associated with some systems showing  quasi-stationary states, and   are the maximizing distributions for the non-additive Tsallis entropy  \cite{tsallis1988possible}:  
\begin{equation}
S_q 
= -k \sum_i p_i^q \ln_q(p_i)
\end{equation}
 where $k$ is a positive constant, $q\neq 1$, 
 and $p_i=p(E_i)$ is the probability that the system is in the $i^{\text{th}}$ configuration and  satisfying $\sum_{i} p_i=1$. The function:
 \begin{equation}
\ln_q(x) =  \frac{x^{(1-q)}-1}{1-q} \quad (x>0)
\end{equation}
 denotes the $q$-logarithm, inverse of the $q$-exponential i.e. $\ln_q[\exp_q(x)]=\exp_q[\ln_q(x)]=x$ \cite{tsallis2004should}.   
 It is clear that as $q\to 1$, $\ln_q(x)\to \ln(x) $, and one recovers the  Boltzmann-Gibbs entropy 
$S_1 = -k_B \sum_{i} p_i \ln (p_i)$ 
 ($k=k_B$ is the Boltzmann constant) associated with the standard exponential and the Gaussian distribution  \cite{tsallis1988possible}.

The $q$-exponential function arises also from the Laplace transform of the the Gamma  probability density function (PDF) \cite{beck2003superstatistics}. 
This means that the EDLC   can be viewed as a distributed network of exponentially decaying $RC$ sub-systems ($e^{-t/\tau}$) weighted by the Gamma PDF.  In other words, the behavior of an EDLC device is equivalent to the collective response of the superposition of a large number of spatially-distributed network of sub-systems, each of them follows the tradition exponential decay \cite{Allagui:2022ue, allagui2021gouy}.

 Alternatively to the modified evolution equation given by Eq.\;\ref{eq3}, from the $q$-difference defined as \cite{borges2004possible}:
\begin{equation}
 x\ominus_q y =\frac{x-y}{1+(1-q)y} \quad (y \neq 1/(q-1))
\end{equation}
 and the $q$-derivative  defined as \cite{borges2004possible}:
\begin{equation}
\mathsf{D}_{(q)} f(x) = \lim_{y\to x} \frac{f(x)-f(y)}{x \ominus_q y}  = [1+(1-q)x] \frac{\mathrm{d}f(x)}{\mathrm{d}x} 
\label{eq26}
\end{equation}
with the corresponding $q$-integral:
\begin{equation}
\int_{(q)} f(x) \mathrm{d}_q x = \int \frac{f(x)}{1+(1-q)x} \mathrm{d}x
\end{equation} 
both under the condition that $[1+(1-q)x] \neq 0$, the result given in Eq.\;\ref{eq4} can also be obtained from the solution of the deformed differential equation:
\begin{equation}
\textcolor{black}{\mathsf{D}_{({q_1})} \rho_{q_1}(t) \equiv [1 - (1-{q_1}) (t/\tau_{q_1})] \frac{\mathrm{d}\rho_{q_1}(t)}{\mathrm{d}t} = -  \frac{\rho_{q_1}(t)}{\tau_{q_1}}}
\end{equation} 
Again at the limit of $q_1 \to 1$ one evidently retrieves  $\rho(t)= e^{- t/\tau}$, otherwise,  {for $q_1<1$, Eq.\;\ref{eq4} belongs to a particular case of a type-1 beta family of functions and  
 for $q_1>1$ (by writing $1-q_1=-(q_1-1)$), Eq.\;\ref{eq4} belongs to a particular case of a type-2 beta family of functions \cite{mathai2012pathway, mathai2005pathway, mathai2007pathway, mathai2015stochastic, sebastian2015overview, jagannathan2020deformed}}.   
The half lifetime characteristic in this case is:
\begin{equation}
t_{1/2} = \frac{\tau_{q_1}}{ 2^{1-{q_1}}} \ln_{q_1}(2)
\end{equation}
which tends to the usual decaying law $\tau \ln(2)$ as   ${q_1}\to 1$. 
%
   
  Finally, we note that from Eq.\;\ref{eq4} with $Z_q(s) = s \mathcal{L}([1- \rho_q(t)];s) =1- s \mathcal{L}(\rho_q(t);s)$ that the following impedance function is obtained:
\begin{equation}
\textcolor{black}{Z_{q_1}(s) = 1-  \eta^{\kappa_1} e^{\eta}  \gamma(1-\kappa_1, \eta)}
\label{eq33}
\end{equation}
     where $\kappa_1=1/({q_1}-1)$ and $\eta=\kappa_1 \tau_{q_1} s$. 

\begin{figure*}[t]
\begin{center}
\includegraphics[height=.275\textwidth]{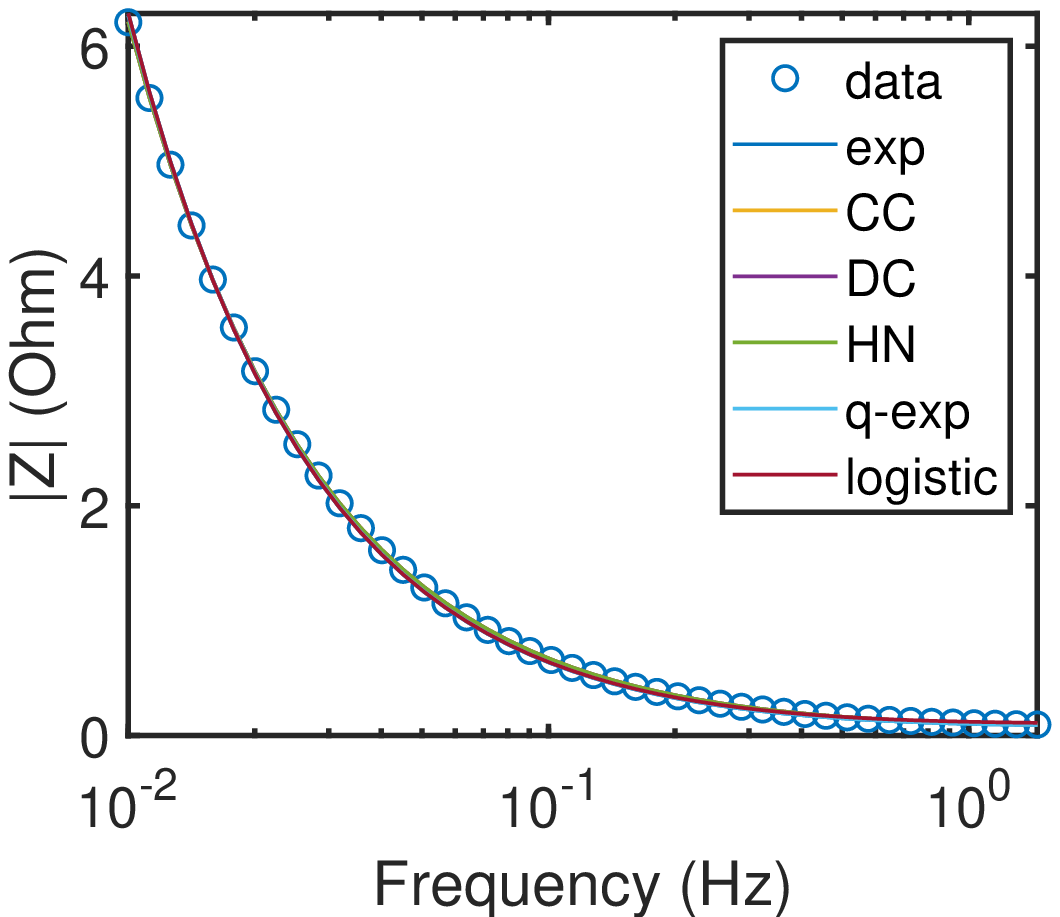}
\includegraphics[height=.275\textwidth]{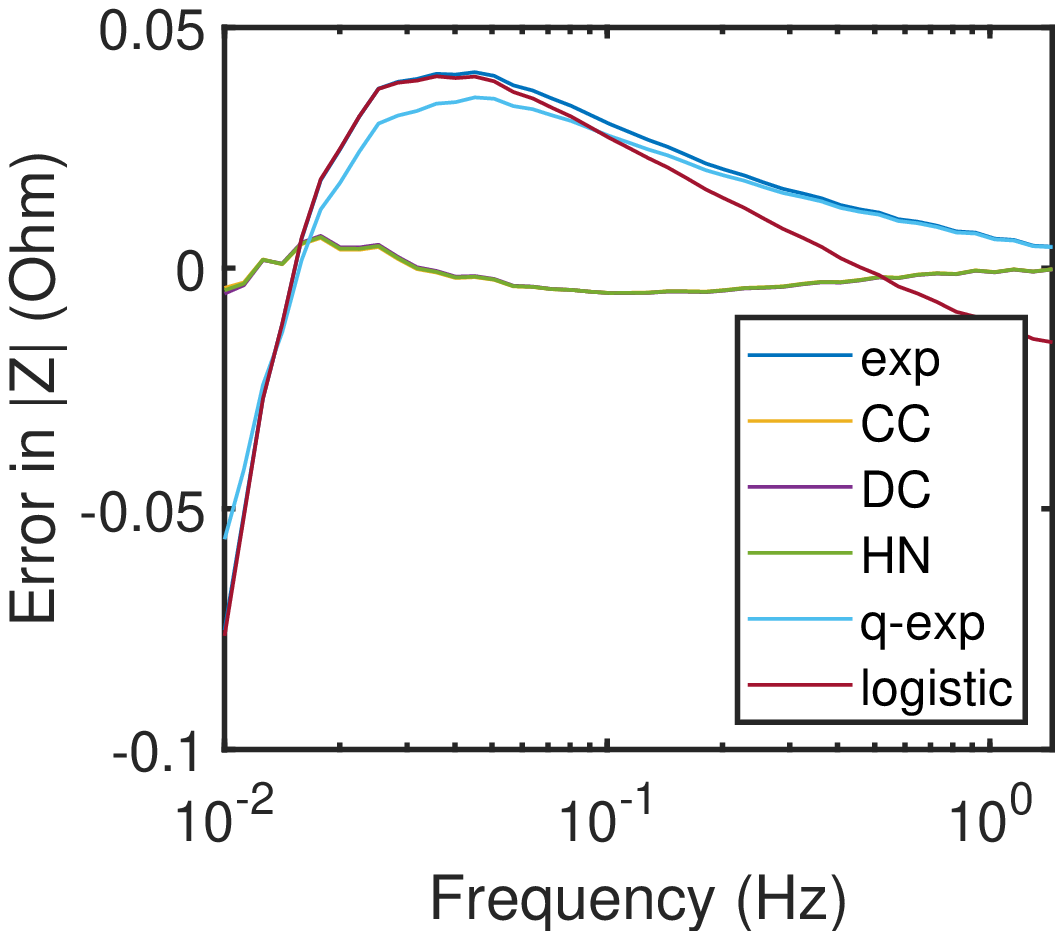}
\includegraphics[height=.275\textwidth]{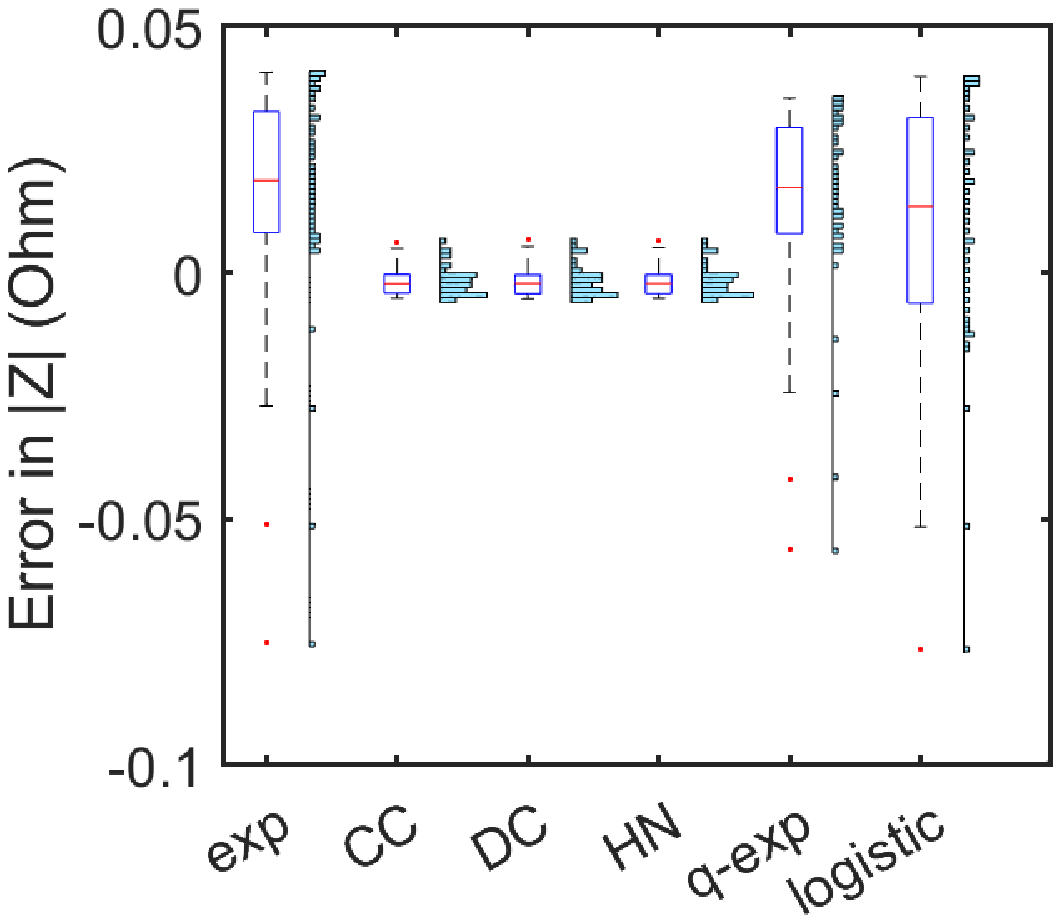} \\  
\includegraphics[height=.275\textwidth]{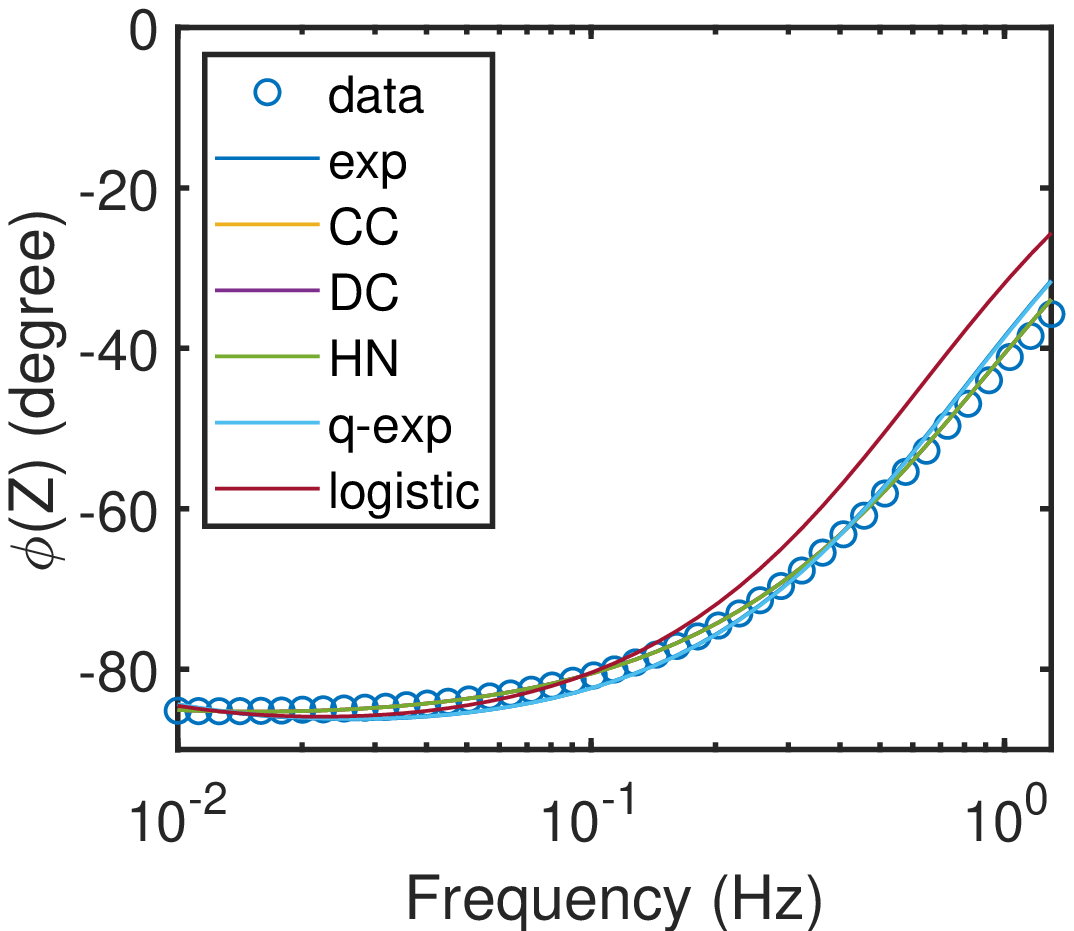}\includegraphics[height=.275\textwidth]{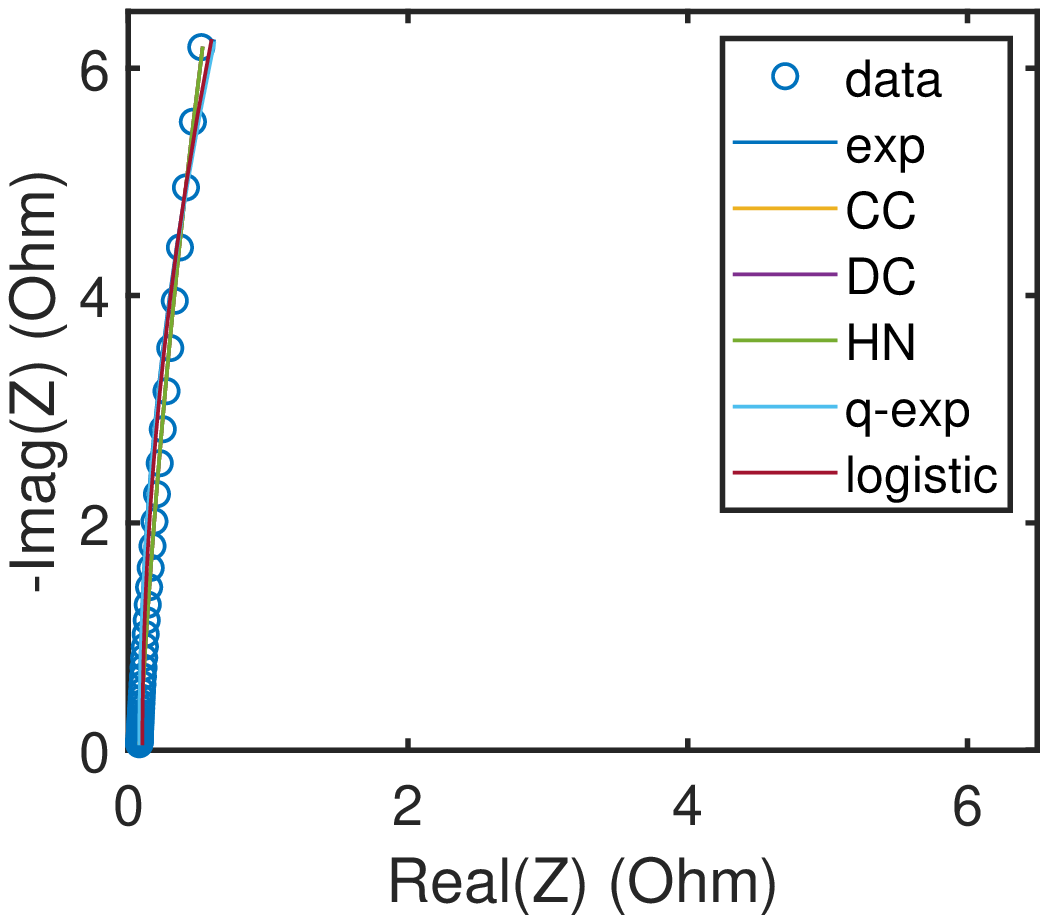}
\caption{Spectral impedance results obtained on a commercial Samxon supercapacitor (part No. DRL105S0TF12RR, rated 2.7\,V, 1\,F) at open-circuit voltage: (a) magnitude of impedance vs. frequency and its corresponding error plots in (b)-(c); (d) impedance phase angle vs. frequency; (e) real vs. imaginary parts of impedance. Fitting results using 
Eq.\;\ref{eq1} ("exp" with $\{R,\tau\}=\{73.60\,\Omega, 185.9\}$ sec),
 Eq.\;\ref{eqCC} ("CC" with $\{R,\tau_{\alpha},\alpha\}=\{209.1\,\Omega, 591.3\,\text{sec}, 0.972\}$), 
 Eq.\;\ref{eqDC} ("DC" with $R=188.6\,\Omega$, $\tau_{\beta}=533.4$ sec, $\beta=0.972$), 
 Eq.\;\ref{eqH} ("HN" with $\{R,\tau_H,\alpha,\beta \}=\{199.9\,\Omega, 565.6\,\text{sec}, 0.984, 0.987\}$), 
 Eq.\;\ref{eq33} ("q-exp" with $\{R,\tau_{q_1},q_1\}=\{5709\,\Omega, 14360\,\text{sec}, q_1=79.84\}$), and 
 Eq.\;\ref{eq39} ("logistic" with $\{R,\tau_{q_2}, q_2\}=\{79.54\,\Omega, 200.9\,\text{sec}, 0.999\}$) are also shown.}
\label{fig1}
\end{center}
\end{figure*}

 On the other hand, the dual derivative operator in Eq.\;\ref{eq26} is defined as \cite{borges2004possible}:
\begin{equation}
\mathsf{D}^{(q)} f(x) =  \lim_{y\to x} \frac{f(x) \ominus_q f(y)}{x - y}  = \frac{1}{[1+(1-q)f(x)]} \frac{\mathrm{d}f(x)}{\mathrm{d}x} 
\end{equation}
with its corresponding $q$-integral:
\begin{equation}
\int^{(q)} f(x) \mathrm{d}_q x = \int [{1+(1-q)x}] {f(x)} \mathrm{d}x
\end{equation} 
 with $  [1+(1-q)f(x)] \neq 0$. Using such a definition, one obtains as a solution for the nonlinear ordinary differential equation:
 \begin{equation}
\textcolor{black}{\mathsf{D}^{({q_2})} \rho_{q_2}(t) \equiv \frac{1}{[1+(1-{q_2}) {\rho_{q_2}(t)}]} \frac{\mathrm{d}\rho_{q_2}(t)}{\mathrm{d}t}  =  -  \frac{\rho_{q_2}(t)}{\tau_{q_2}}}
\end{equation}
the following result in terms of the logistic function:
\begin{equation}
\textcolor{black}{\rho_{q_2}(t)   = \frac{1}{({q_2}-1)+ (2-{q_2})  e^{t/\tau_{q_2}}   }}
\label{eq31}
\end{equation}
It is evident that at the limit $q_2\to 1$, one retrieves the traditional exponential decay given by Eq.\;\ref{eq2}.  
The half lifetime is 
\begin{equation}
t_{1/2} = \tau_{q_2} \ln \left( \frac{3-{q_2}}{2-{q_2}} \right)
\end{equation}
which reduces as expected to $ \tau_D  \ln(2)$ as ${q_2}\to 1$. 
The impedance function corresponding to  Eq.\;\ref{eq31} is:
\begin{equation}
\textcolor{black}{Z_{q_2}(s) = 1+ \frac{ \tau_{q_2} s \;{}_2F_{1}(1,1+\tau_{q_2} s,2+ \tau_{q_2} s, {(1-\kappa_2})^{-1}) }{({q_2}-2) (1+ \tau_{q_2} s)}
\label{eq39}
}
\end{equation}
where ${}_2F_{1}(a,b,c,z)$ is the hypergeometric function and       $\kappa_2=1/({q_2}-1)$. It simplifies to $({1+s \tau_D})^{-1}$ at the limit of ${q_2}=1$.

\begin{figure*}[t]
\begin{center}
\includegraphics[height=.275\textwidth]{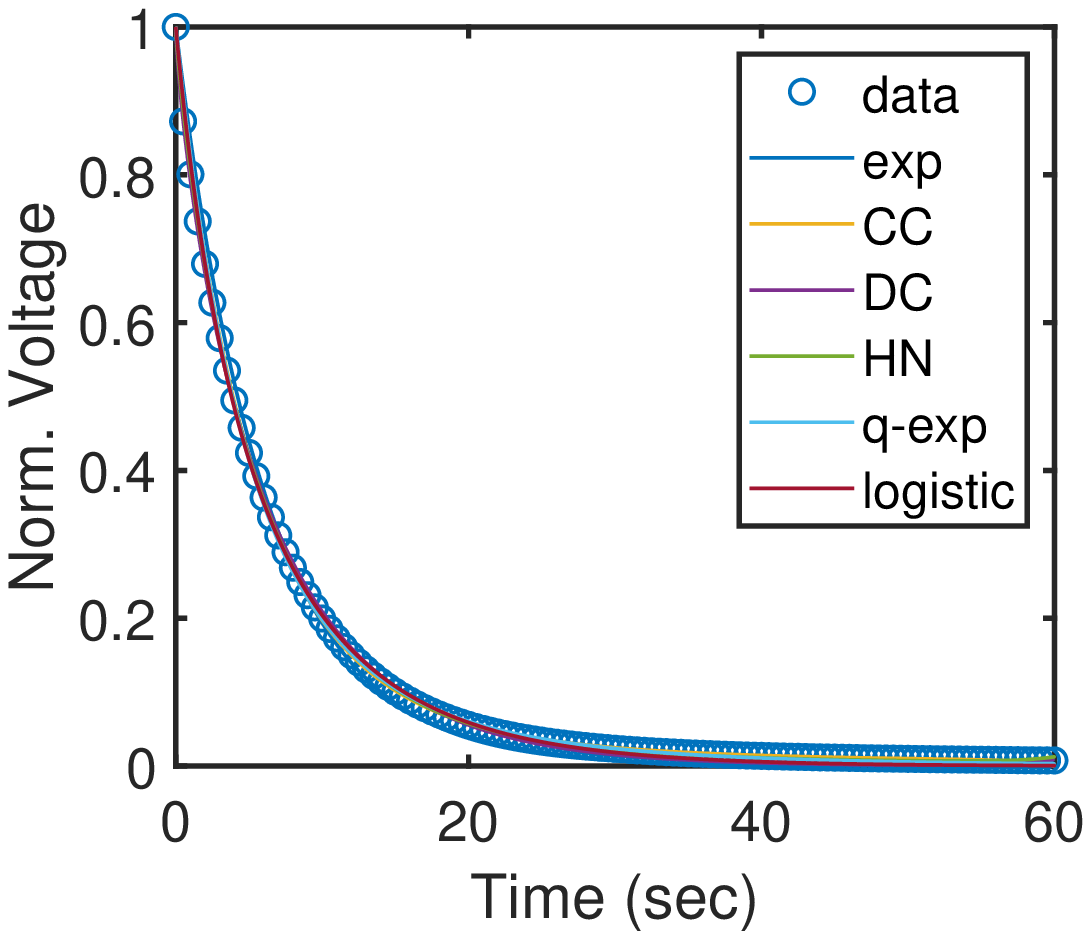}
\includegraphics[height=.275\textwidth]{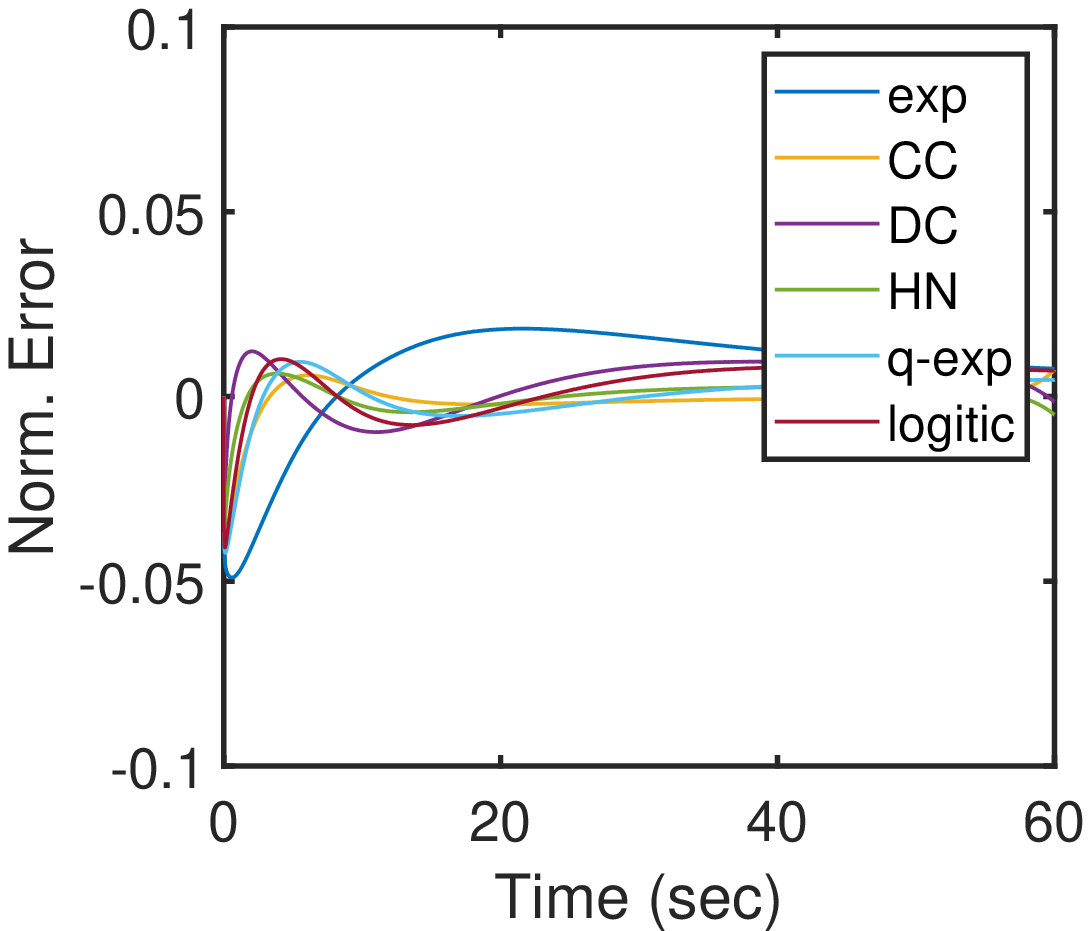}
\includegraphics[height=.275\textwidth]{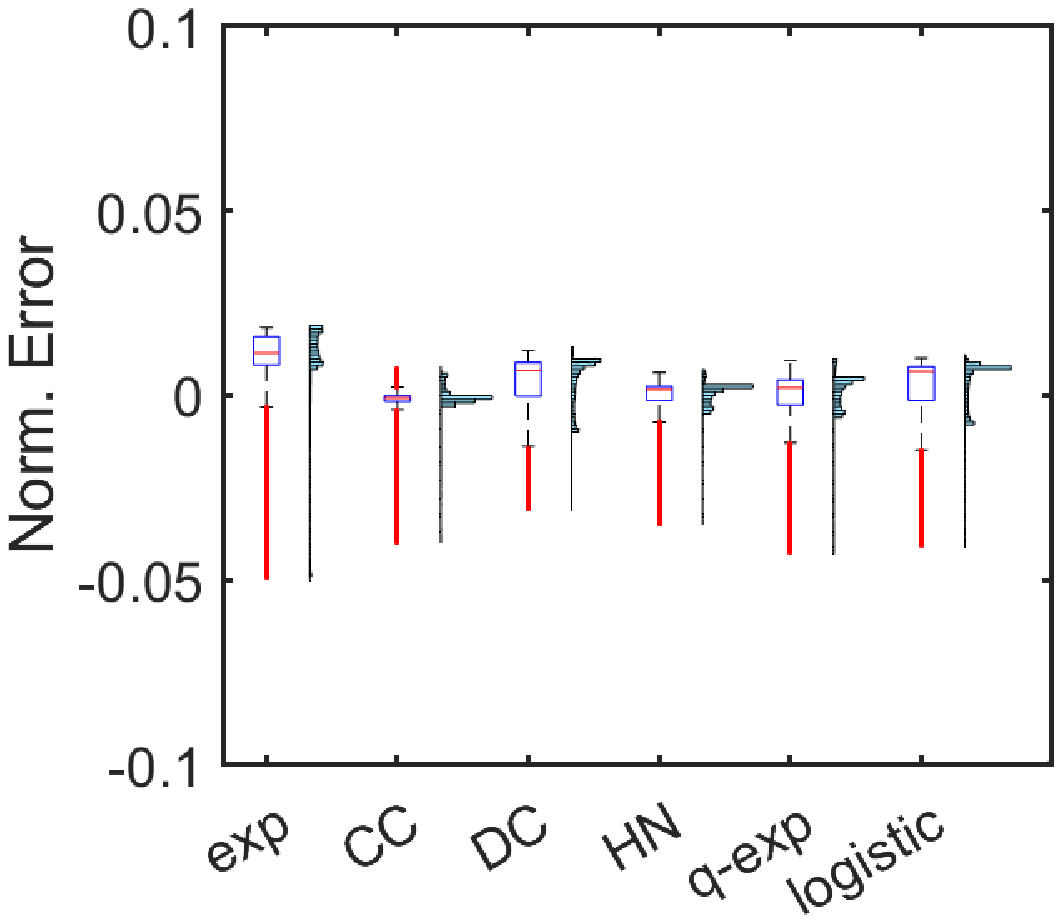} \\  
\includegraphics[height=.275\textwidth]{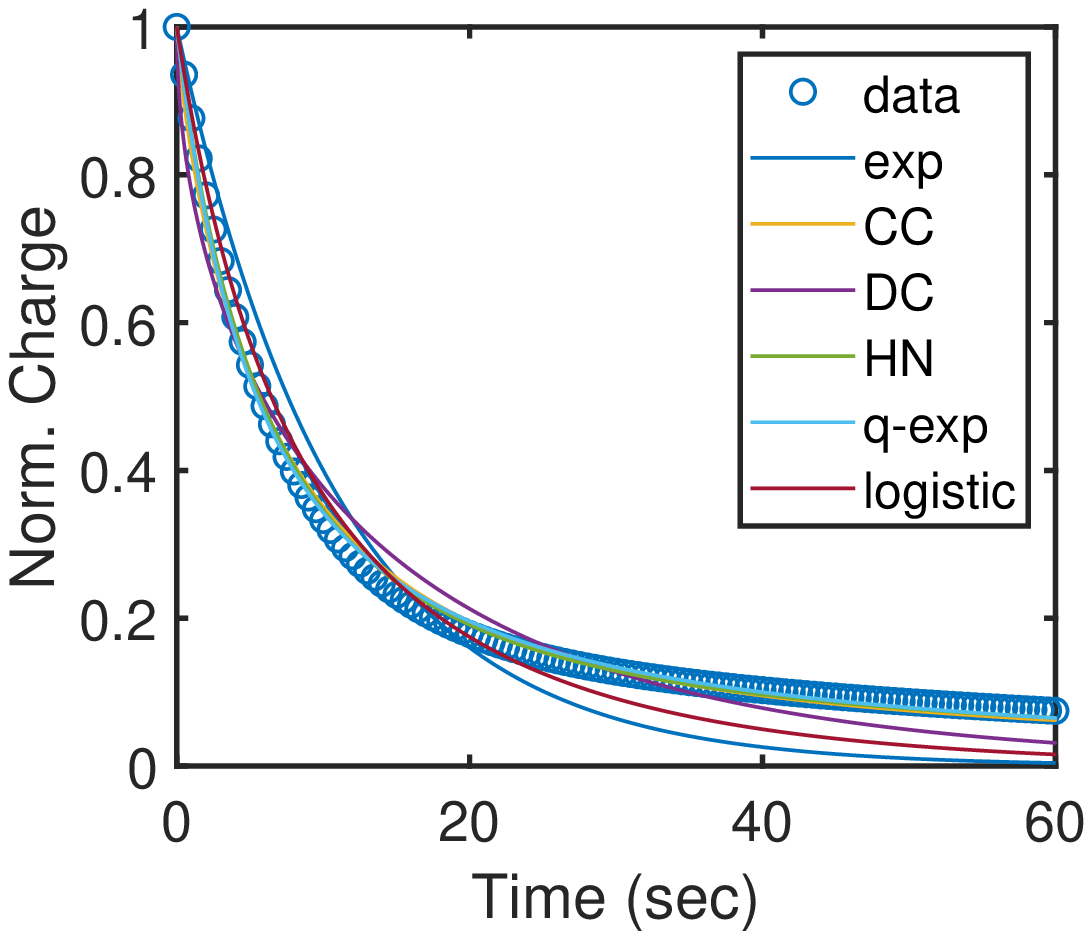}\includegraphics[height=.275\textwidth]{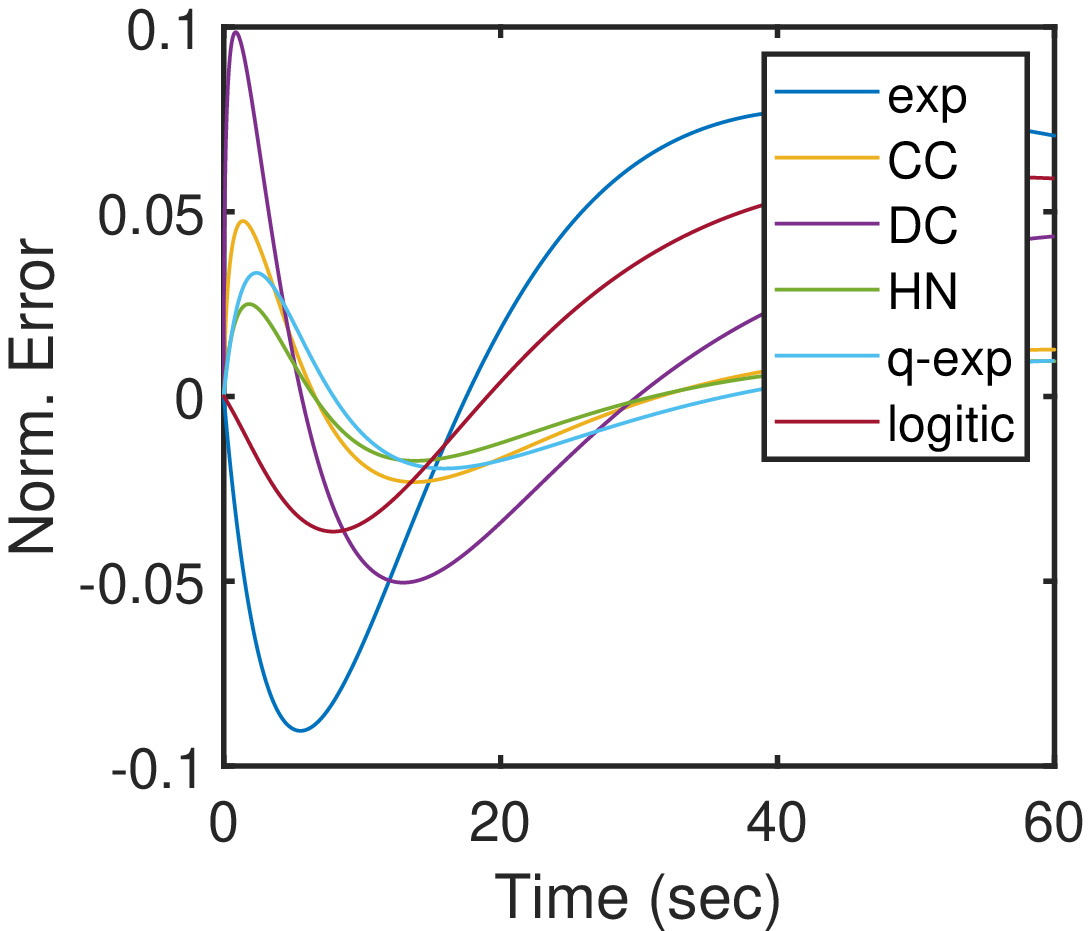}\includegraphics[height=.275\textwidth]{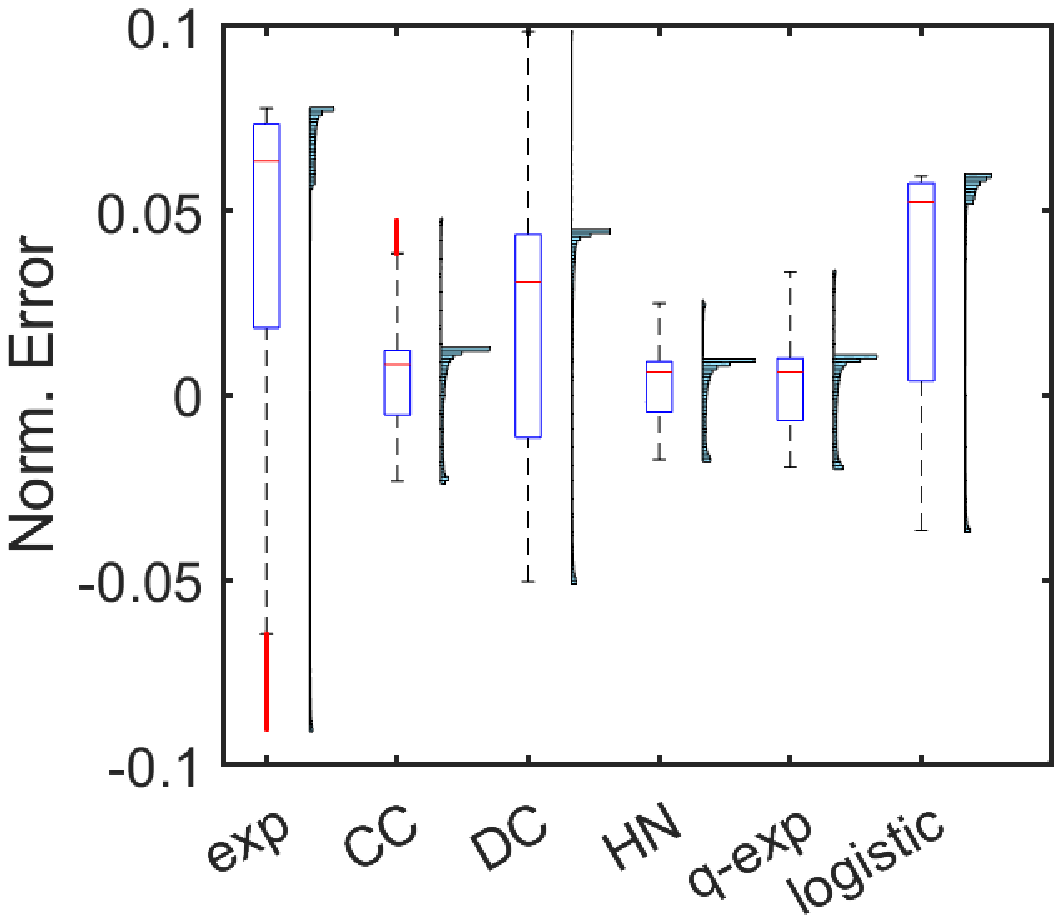}
\caption{Plots of (a) normalized voltage and (d) normalized charge vs. time measured on a pre-charged commercial Samxon supercapacitor (part No. DRL105S0TF12RR, rated 2.7\,V, 1\,F) when connected to a resistive load of 2\,Ohm. Fitting results for voltage/charge respectively using 
Eq.\;\ref{eq2} 
("exp" with 
$\tau=\{6.092,10.92\}$\,\text{sec}),  
Eq.\;\ref{eq41} 
("CC" with 
$\{\tau_{\alpha},\alpha\}=\{5.750\,\text{sec}, 0.946\}$, $\{8.698\,\text{sec},0.800\}$),  
Eq.\;\ref{eq12} 
("DC" with 
$\{\tau_{\beta},\beta\}=\{8.296\,\text{sec},0.760\}$, $\{26.00\,\text{sec}, 0.496\}$), 
 Eq.\;\ref{eqHH} 
 ("HN" with 
 $\{\tau_H,\alpha,\beta\}=\{6.709\,\text{sec},0.964,{0.888}\}$, $\{8.689\,\text{sec},0.802, {1.000}\}$), 
 Eq.\;\ref{eq4} 
 ("$q$-exp" with 
 $\{\tau_{q_1},q_1\}=\{5.144\,\text{sec}, 1.221\}$, $\{5.829\,\text{sec},1.822\}$) 
  and  
  Eq.\;\ref{eq31} ("logistic" with 
  $\{\tau_{q_2},q_2\}=\{8.843\,\text{sec}, 0.141\}$, $\{18.00\,\text{sec},-0.319\}$) are also shown.   Normalized errors plotted vs. time (b,e) and in box plot format (c,f) for the different models are presented for  voltage and charge respectively.}
\label{fig2}
\end{center}
\end{figure*}

\section{Experimental Results\label{RD}}

The electrical measurements were carried out on a commercial Samxon supercapacitor, part No. DRL105S0TF12RR, rated 2.7\,V, 1\,F using a Biologic VSP-300 electrochemical station equipped with impedance spectroscopy module. 
The  spectral impedance results were obtained at open-circuit voltage with stepped sine excitations from 100\,kHz to 10\,mHz frequency and with 10\,mV rms amplitude.  
The time-domain charge and discharge measurements were conducted as follows. First, the device was pre-charged with constant current-constant voltage (CCCV) mode: 100\,mA up to the nominal voltage of 2.7\,V, and then the voltage was maintained at 2.7\,V for 5 minutes. For the subsequent discharge step, the potentiostat acted as a constant resistor $R$ (2, 10, 50\,Ohm) for the duration necessary for the voltage to drop from 2.7\,V to 3\,mV.

\subsection{Impedance spectroscopy}

 Plot of magnitude of impedance vs. frequency are shown in Fig.\;\ref{fig1}(a). We limited the frequency range to $[0.01;1.41]$\,Hz where the device is mostly capacitive. Superposed on the experimental data are   fitting by the metaheuristic particle swarm optimization (PSO) technique \cite{jordehi2016time} using the six models given in Eq.\;\ref{eq1} (denoted "exp" in the legend),
 Eq.\;\ref{eqCC} ("CC"), 
 Eq.\;\ref{eqDC} ("DC"), 
 Eq.\;\ref{eqH} ("HN"), 
 Eq.\;\ref{eq33} ("q-exp"), and 
 Eq.\;\ref{eq39} ("logistic").  The models parameters are given in the figure caption. 
 In Figs.\;\ref{fig1}(b)-(c), we show the corresponding errors in magnitude of impedance vs. frequency and in box plot format, respectively. Fig.\;\ref{fig1}(d) depicts the experimental impedance phase angle vs. frequency and fitting using the six models presented in this study. The same is shown in Fig.\;\ref{fig1}(e)
 for the real vs. imaginary parts of impedance   with the frequency being an implicit variable.  The deviation from the -90 deg. phase angle in Fig.\;\ref{fig1}(d) and the non-vertical profiles in Fig.\;\ref{fig1}(d)(e) indicate the non-ideal capacitive-resistive behavior of the device. Detailed analysis of spectral impedance of EDLCs can be found for instance in ref. \cite{eis}.  
 
 It is clear from these results that fitting with the fractional-order models outperform those done with the $q$-deformed models. The root mean square errors (RMSE) over the  frequency range  $[0.01;1.41]$\,Hz are 
 0.042, 
0.005, 
0.005, 
0.005, 
0.038, and
0.037 for Eq.\;\ref{eq1},
 Eq.\;\ref{eqCC}, 
 Eq.\;\ref{eqDC}, 
 Eq.\;\ref{eqH}, 
 Eq.\;\ref{eq33}, and 
 Eq.\;\ref{eq39}, respectively. 
 Furthermore, an important advantage of the fractional-order impedance models is the simplicity of the expressions which makes their computation much more efficient than the new proposed $q$-deformed models. The latter ones involve computational weight with the numerical estimation of the incomplete gamma function (Eq.\;\ref{eq33}) or the infinite sum in the hypergeometric function (Eq.\;\ref{eq39}). 
 Nonetheless, the errors associated with the $q$-deformed models  are still comparatively reasonable and slightly better that the classical $RC$ model. 

\subsection{Discharge response}

 The results for the first 60 sec of the discharge relaxation sequence for the case of $R=\text{2\,Ohm}$ are shown in Fig.\;\ref{fig2}(a) in terms of normalized voltage vs. time and in Fig.\;\ref{fig2}(d) for the normalized charge vs. time. The resistive load of  $R=\text{2\,Ohm}$  is taken  as an illustrative example as the same conclusions can be drawn from the other trials (not shown here). The fitting results  (also carried out by PSO technique) using the models given by Eq.\;\ref{eq2} (denoted "exp" in the legend), Eq.\;\ref{eq41} ("CC"),  Eq.\;\ref{eq12} ("DC"), Eq.\;\ref{eqHH} ("HN"), Eq.\;\ref{eq4} ("q-exp") and Eq.\;\ref{eq31} ("logistic") are also shown.  
The models parameters are   given in the figure caption. 
We also plootted the normalized error vs. time and in  box plot format  for the different models.  
 
 It is clear from the figure that the exponential decay model (Eq.\;\ref{eq2}) is the least efficient  in properly capturing  the experimental data of the non-ideal, dissipative EDLC device. The RMSE for the (normalized) voltage and charge fitting are found to be 0.016 and 0.065, respectively. We also remark that the fitting time constant $\tau$ for the  voltage is 6.092 sec whereas for the charge it is 10.92 sec. This indicates  that the relation $q(t)=C v(t)$ where $C$ is a constant capacitance is not valid for non-ideal capacitive devices \cite{ieeeted, allagui2021inverse,acs2,QV}. Fractional-order and $q$-deformed exponential models, on the other hand,  whether they are with one or more extra degrees of freedom are evidently more appropriate at closely following the non-exponential data. 
 The RMSE associated with the models of Eq.\;\ref{eq41},  Eq.\;\ref{eq12}, Eq.\;\ref{eqHH}, Eq.\;\ref{eq4} and Eq.\;\ref{eq31}
 are 0.005, 
0.007, 
0.004, 
0.006, 
0.008, respectively for voltage fitting and 
0.015, 
0.039, 
0.015, 
0.012, 
0.046, respectively for charge fitting. 
 In particular, for both voltage  and charge  profiles, it seems like the Cole-Cole-based model (Eq.\;\ref{eq41}) and its generalization to the Havriliak-Negami model (Eq.\;\ref{eqHH}) are the best performing fractional-order models for this case. The $q$-exponential model (Eq.\;\ref{eq4}) with $q\neq 1$ has proved to be equally reliable too as indicated from the error of regressions and their plots.

 However,  the evolution kinetic equation for $\rho(t)$ given by Eq.\;\ref{eq3} in which the local time derivative of $\rho(t)$ is proportional to a power of $\rho(t)$ is physically more tractable than the fractional-order equations\;\ref{eq217} or\;\ref{eqRL1} for the Cole-Cole model or Eq.\;\ref{eqEEqHN} for the Havriliak-Negami model. As illustrated schematically in Fig.\;\ref{fig1},  the non-local (global) integro-differential fractional models  require knowledge of all events over the history of the device in order to estimate its current state at an instant $t$ \cite{memoryAPL, memQ, allagui2021possibility, nigmatullin1984theory}. Even though the long-term memory effect tends to fade away quickly compared to the short term, it is still a challenging task to gather all prior state information  about a device described with blackfractional-order models. It is worth mentioning again that the   $q$-exponential function is directly connected to the non-additive Tsallis entropy  \cite{tsallis1988possible} and to Beck and Cohen's superstatistics \cite{beck2003superstatistics}, which  makes its physical interpretation easier to justify. 
 Finally, we note that the $q$-exponential function has a simple algebraic form, and thus more     efficient to compute compared to the cumbersome series associated with the Mittag-Leffler and the $H$-functions that appear in the solutions of fractional-order equations.  
 This is the opposite of what we have seen with the fitting models of the impedance data. The fractional impedance models are simple fractions with one or two power coefficients, whereas the $q$-deformed models involving the  incomplete gamma function or the hypergeometric series are numerically more demanding to compute.

\section{Conclusion}

In this work we reviewed different ways of generalizing the evolution equation given by Eq.\;\ref{eqrho} by first replacing the first-order time derivative $\mathrm{d}\rho(t)/\mathrm{d}t$ with different forms of fractional-order derivatives leading to the well-known Cole-Cole, Davidson-Cole and Havriliak–Negami models, both in the frequency domain and their corresponding relaxation functions (for charge and voltage) in the time domain. We then proposed a new model in which   $\rho(t)$ in Eq.\;\ref{eqrho} is replaced by a power function, i.e. $[\rho(t)]^{q} (q \in \mathbb{R})$, which can also be expressed in terms of a linear $q$-deformed derivative. This led to a relaxation function in terms of $q$-exponential function able to capture quite efficiently the nonexponential behavior observed with a commercial non-ideal (dissipative) EDLC device. The dual nonlinear $q$-derivative led to a logistic function-type  of   decay. The $q$-deformed impedance functions involving the incomplete gamma function or the hypergeometric function are also fitted to the experimental data and compared with the traditional fractional-order models. All non-Debye models, given their extra degrees of freedom, are capable of capturing the time-domain and frequency-domain data with great fidelity. However, the fractional-order models are found to be computationally more efficient with the impedance fitting, whereas the $q$-deformed models are more efficient with the time-domain relaxation data. In addition, as we noted the $q$-deformed evolution equation maintains the first-order derivative of the evolving function, and is directly  associated with the Tsallis thermostatics and Beck and Cohen superstatistics, which makes its  physical comprehension more straightforward  that the non-local fractional-order integro-differential evolution  equations. 

\section*{Authors' Contributions}

A.A. developed the idea behind this work. 
A.A. and H.B. derived the mathematical expressions. 
A.A. designed and carried out the experimental measurements. 
M.A. carried out the data fitting.
A.A. and A.S.E. discussed and analyzed the results. 
A.A. wrote the manuscript. All co-authors reviewed the manuscript and provided feedback.

\section*{Data Availability}
The data that support the findings of this study are available
from the corresponding author upon reasonable request.

\appendix

\section{Fox's $H$-function\label{H}}

Fox's $H$-function 
 is defined in terms of  Mellin-Barnes type  integral as \cite{mathai2009h}:
\begin{align}
H^{m,n}_{p,q}(z) &= H^{m,n}_{p,q}\left[ z|^{(a_p,A_p)}_{(b_q,B_q)} \right] \nonumber \\
&=H^{m,n}_{p,q}\left[ z|^{(a_1,A_1),\ldots,(a_p,A_p)}_{(b_1,B_1),\ldots, (b_q,B_q)} \right] \nonumber \\
&=\frac{1}{2\pi i} \int_L h(s) z^{-s} ds
\end{align}
 where $h(s)$ is given by:
 \begin{equation}
h(s) = \frac{\left\{\prod_{j=1}^m \Gamma(b_j + B_j s)\right\}  \left\{\prod_{j=1}^n \Gamma(1-a_j - A_j s)\right\}}
{\left\{\prod_{j={m+1}}^q \Gamma(1-b_j - B_j s)\right\} \left\{\prod_{j={n+1}}^p \Gamma(a_j + A_j s)\right\}}
\end{equation}
where $i=\sqrt{-1}$, 
$m,n,p,q$ are  integers satisfying ($0 \leqslant n \leqslant p$, $1 \leqslant m \leqslant q$), 
$z\neq 0$, and $z^{-s}=\exp \left[ -s (\ln|z|+ i \arg z) \right] $,  $A_i, B_j \in \mathbb{R}_+$, $a_i, b_j \in \mathbb{R}$ or $\mathbb{C}$ with $(i=1,\ldots,p)$, $(j=1,\ldots,q)$. 
The contour of integration $L$ is a suitable contour separating the poles $-(b_j+\nu)/B_j$, ($j=1,\ldots,m$; $\nu=0, 1, 2, \ldots$),  of the gamma functions $\Gamma(b_j+ B_j s)$ from the poles $(1-a_{\lambda} +k)/A_{\lambda}$, ($\lambda=1,\ldots,n$; $k=0, 1, 2, \ldots$) of the gamma functions  $\Gamma (1-a_{\lambda} - A_{\lambda} s)$, that is $A_{\lambda} (b_j+ \nu) \neq B_j (a_{\lambda - k - 1})$. 
 An empty product is always interpreted as unity. 
A  comprehensive account of the $H$-function is available in Mathai,  Saxena, and   Haubold \cite{mathai2009h}
 and Mathai and  Saxena \cite{mathai1978h}

\section*{References}


\begin{thebibliography}{47}%
\makeatletter
\providecommand \@ifxundefined [1]{%
 \@ifx{#1\undefined}
}%
\providecommand \@ifnum [1]{%
 \ifnum #1\expandafter \@firstoftwo
 \else \expandafter \@secondoftwo
 \fi
}%
\providecommand \@ifx [1]{%
 \ifx #1\expandafter \@firstoftwo
 \else \expandafter \@secondoftwo
 \fi
}%
\providecommand \natexlab [1]{#1}%
\providecommand \enquote  [1]{``#1''}%
\providecommand \bibnamefont  [1]{#1}%
\providecommand \bibfnamefont [1]{#1}%
\providecommand \citenamefont [1]{#1}%
\providecommand \href@noop [0]{\@secondoftwo}%
\providecommand \href [0]{\begingroup \@sanitize@url \@href}%
\providecommand \@href[1]{\@@startlink{#1}\@@href}%
\providecommand \@@href[1]{\endgroup#1\@@endlink}%
\providecommand \@sanitize@url [0]{\catcode `\\12\catcode `\$12\catcode
  `\&12\catcode `\#12\catcode `\^12\catcode `\_12\catcode `\%12\relax}%
\providecommand \@@startlink[1]{}%
\providecommand \@@endlink[0]{}%
\providecommand \url  [0]{\begingroup\@sanitize@url \@url }%
\providecommand \@url [1]{\endgroup\@href {#1}{\urlprefix }}%
\providecommand \urlprefix  [0]{URL }%
\providecommand \Eprint [0]{\href }%
\providecommand \doibase [0]{https://doi.org/}%
\providecommand \selectlanguage [0]{\@gobble}%
\providecommand \bibinfo  [0]{\@secondoftwo}%
\providecommand \bibfield  [0]{\@secondoftwo}%
\providecommand \translation [1]{[#1]}%
\providecommand \BibitemOpen [0]{}%
\providecommand \bibitemStop [0]{}%
\providecommand \bibitemNoStop [0]{.\EOS\space}%
\providecommand \EOS [0]{\spacefactor3000\relax}%
\providecommand \BibitemShut  [1]{\csname bibitem#1\endcsname}%
\let\auto@bib@innerbib\@empty
\bibitem [{\citenamefont {Zhao}\ and\ \citenamefont
  {Burke}(2021)}]{zhao2021electrochemical}%
  \BibitemOpen
  \bibfield  {author} {\bibinfo {author} {\bibfnamefont {J.}~\bibnamefont
  {Zhao}}\ and\ \bibinfo {author} {\bibfnamefont {A.~F.}\ \bibnamefont
  {Burke}},\ }\bibfield  {title} {\enquote {\bibinfo {title} {Electrochemical
  capacitors: materials, technologies and performance},}\ }\href@noop {}
  {\bibfield  {journal} {\bibinfo  {journal} {Energy Storage Mater.}\ }\textbf
  {\bibinfo {volume} {36}},\ \bibinfo {pages} {31--55} (\bibinfo {year}
  {2021})}\BibitemShut {NoStop}%
\bibitem [{\citenamefont {Zhao}\ \emph {et~al.}(2021)\citenamefont {Zhao},
  \citenamefont {Taylor}, \citenamefont {Hu}, \citenamefont {Evanko},
  \citenamefont {Zeng}, \citenamefont {Wang}, \citenamefont {Ohnishi},
  \citenamefont {Tsukazaki}, \citenamefont {Li}, \citenamefont {Stadie} \emph
  {et~al.}}]{zhao2021structural}%
  \BibitemOpen
  \bibfield  {author} {\bibinfo {author} {\bibfnamefont {Y.}~\bibnamefont
  {Zhao}}, \bibinfo {author} {\bibfnamefont {E.~E.}\ \bibnamefont {Taylor}},
  \bibinfo {author} {\bibfnamefont {X.}~\bibnamefont {Hu}}, \bibinfo {author}
  {\bibfnamefont {B.}~\bibnamefont {Evanko}}, \bibinfo {author} {\bibfnamefont
  {X.}~\bibnamefont {Zeng}}, \bibinfo {author} {\bibfnamefont {H.}~\bibnamefont
  {Wang}}, \bibinfo {author} {\bibfnamefont {R.}~\bibnamefont {Ohnishi}},
  \bibinfo {author} {\bibfnamefont {T.}~\bibnamefont {Tsukazaki}}, \bibinfo
  {author} {\bibfnamefont {J.-F.}\ \bibnamefont {Li}}, \bibinfo {author}
  {\bibfnamefont {N.~P.}\ \bibnamefont {Stadie}}, \emph {et~al.},\ }\bibfield
  {title} {\enquote {\bibinfo {title} {What structural features make porous
  carbons work for redox-enhanced electrochemical capacitors? a fundamental
  investigation},}\ }\href@noop {} {\bibfield  {journal} {\bibinfo  {journal}
  {ACS Energy Lett.}\ }\textbf {\bibinfo {volume} {6}},\ \bibinfo {pages}
  {854--861} (\bibinfo {year} {2021})}\BibitemShut {NoStop}%
\bibitem [{\citenamefont {Yang}\ \emph {et~al.}(2022)\citenamefont {Yang},
  \citenamefont {Jia}, \citenamefont {Li}, \citenamefont {Yang}, \citenamefont
  {Zhang}, \citenamefont {Huang}, \citenamefont {Zheng}, \citenamefont {Li},\
  and\ \citenamefont {Shao}}]{yang2022understanding}%
  \BibitemOpen
  \bibfield  {author} {\bibinfo {author} {\bibfnamefont {Q.}~\bibnamefont
  {Yang}}, \bibinfo {author} {\bibfnamefont {X.}~\bibnamefont {Jia}}, \bibinfo
  {author} {\bibfnamefont {X.}~\bibnamefont {Li}}, \bibinfo {author}
  {\bibfnamefont {Q.}~\bibnamefont {Yang}}, \bibinfo {author} {\bibfnamefont
  {T.}~\bibnamefont {Zhang}}, \bibinfo {author} {\bibfnamefont
  {X.}~\bibnamefont {Huang}}, \bibinfo {author} {\bibfnamefont
  {Q.}~\bibnamefont {Zheng}}, \bibinfo {author} {\bibfnamefont
  {C.}~\bibnamefont {Li}},\ and\ \bibinfo {author} {\bibfnamefont
  {J.}~\bibnamefont {Shao}},\ }\bibfield  {title} {\enquote {\bibinfo {title}
  {Understanding the capacitive charge in bulk porous electrodes by
  mathematical modeling},}\ }\href@noop {} {\bibfield  {journal} {\bibinfo
  {journal} {Phys. Rev. Appl.}\ }\textbf {\bibinfo {volume} {17}},\ \bibinfo
  {pages} {044045} (\bibinfo {year} {2022})}\BibitemShut {NoStop}%
\bibitem [{\citenamefont {Allagui}\ \emph {et~al.}(2018)\citenamefont
  {Allagui}, \citenamefont {Freeborn}, \citenamefont {Elwakil}, \citenamefont
  {Fouda}, \citenamefont {Maundy}, \citenamefont {Radwanh}, \citenamefont
  {Said},\ and\ \citenamefont {Abdelkareem}}]{fracorderreview}%
  \BibitemOpen
  \bibfield  {author} {\bibinfo {author} {\bibfnamefont {A.}~\bibnamefont
  {Allagui}}, \bibinfo {author} {\bibfnamefont {T.~J.}\ \bibnamefont
  {Freeborn}}, \bibinfo {author} {\bibfnamefont {A.~S.}\ \bibnamefont
  {Elwakil}}, \bibinfo {author} {\bibfnamefont {M.~E.}\ \bibnamefont {Fouda}},
  \bibinfo {author} {\bibfnamefont {B.~J.}\ \bibnamefont {Maundy}}, \bibinfo
  {author} {\bibfnamefont {A.~G.}\ \bibnamefont {Radwanh}}, \bibinfo {author}
  {\bibfnamefont {Z.}~\bibnamefont {Said}},\ and\ \bibinfo {author}
  {\bibfnamefont {M.~A.}\ \bibnamefont {Abdelkareem}},\ }\bibfield  {title}
  {\enquote {\bibinfo {title} {Review of fractional-order electrical
  characterization of supercapacitors},}\ }\href@noop {} {\bibfield  {journal}
  {\bibinfo  {journal} {J. Power Sources}\ }\textbf {\bibinfo {volume} {400}}
  (\bibinfo {year} {2018})}\BibitemShut {NoStop}%
\bibitem [{\citenamefont {Allagui}, \citenamefont {Elwakil},\ and\
  \citenamefont {Fouda}(2021)}]{ieeeted}%
  \BibitemOpen
  \bibfield  {author} {\bibinfo {author} {\bibfnamefont {A.}~\bibnamefont
  {Allagui}}, \bibinfo {author} {\bibfnamefont {A.~S.}\ \bibnamefont
  {Elwakil}},\ and\ \bibinfo {author} {\bibfnamefont {M.~E.}\ \bibnamefont
  {Fouda}},\ }\bibfield  {title} {\enquote {\bibinfo {title} {Revisiting the
  time-domain and frequency-domain definitions of capacitance},}\ }\href@noop
  {} {\bibfield  {journal} {\bibinfo  {journal} {IEEE Trans. Electron Devices}\
  }\textbf {\bibinfo {volume} {68}} (\bibinfo {year} {2021})}\BibitemShut
  {NoStop}%
\bibitem [{\citenamefont {Allagui}\ and\ \citenamefont
  {Fouda}(2021)}]{allagui2021inverse}%
  \BibitemOpen
  \bibfield  {author} {\bibinfo {author} {\bibfnamefont {A.}~\bibnamefont
  {Allagui}}\ and\ \bibinfo {author} {\bibfnamefont {M.~E.}\ \bibnamefont
  {Fouda}},\ }\bibfield  {title} {\enquote {\bibinfo {title} {Inverse problem
  of reconstructing the capacitance of electric double-layer capacitors},}\
  }\href@noop {} {\bibfield  {journal} {\bibinfo  {journal} {Electrochim.
  Acta}\ ,\ \bibinfo {pages} {138848}} (\bibinfo {year} {2021})}\BibitemShut
  {NoStop}%
\bibitem [{\citenamefont {Allagui}, \citenamefont {Elwakil},\ and\
  \citenamefont {Eleuch}(2021)}]{acs2}%
  \BibitemOpen
  \bibfield  {author} {\bibinfo {author} {\bibfnamefont {A.}~\bibnamefont
  {Allagui}}, \bibinfo {author} {\bibfnamefont {A.~S.}\ \bibnamefont
  {Elwakil}},\ and\ \bibinfo {author} {\bibfnamefont {H.}~\bibnamefont
  {Eleuch}},\ }\bibfield  {title} {\enquote {\bibinfo {title} {Highlighting a
  common confusion in the computation of capacitance of electrochemical energy
  storage devices},}\ }\href@noop {} {\bibfield  {journal} {\bibinfo  {journal}
  {J. Phys. Chem. C}\ }\textbf {\bibinfo {volume} {125}},\ \bibinfo {pages}
  {9591--9592} (\bibinfo {year} {2021})}\BibitemShut {NoStop}%
\bibitem [{\citenamefont {Allagui}\ and\ \citenamefont
  {Benaoum}(2022)}]{10.1149/1945-7111/ac621e}%
  \BibitemOpen
  \bibfield  {author} {\bibinfo {author} {\bibfnamefont {A.}~\bibnamefont
  {Allagui}}\ and\ \bibinfo {author} {\bibfnamefont {H.}~\bibnamefont
  {Benaoum}},\ }\bibfield  {title} {\enquote {\bibinfo {title} {Power-law
  charge relaxation of inhomogeneous porous capacitive electrodes},}\
  }\href@noop {} {\bibfield  {journal} {\bibinfo  {journal} {J. Electrochem.
  Soc.}\ }\textbf {\bibinfo {volume} {169}},\ \bibinfo {pages} {040509}
  (\bibinfo {year} {2022})}\BibitemShut {NoStop}%
\bibitem [{\citenamefont {Allagui}\ \emph {et~al.}(2016)\citenamefont
  {Allagui}, \citenamefont {Elwakil}, \citenamefont {Maundy},\ and\
  \citenamefont {Freeborn}}]{eis}%
  \BibitemOpen
  \bibfield  {author} {\bibinfo {author} {\bibfnamefont {A.}~\bibnamefont
  {Allagui}}, \bibinfo {author} {\bibfnamefont {A.~S.}\ \bibnamefont
  {Elwakil}}, \bibinfo {author} {\bibfnamefont {B.~J.}\ \bibnamefont
  {Maundy}},\ and\ \bibinfo {author} {\bibfnamefont {T.~J.}\ \bibnamefont
  {Freeborn}},\ }\bibfield  {title} {\enquote {\bibinfo {title} {Spectral
  capacitance of series and parallel combinations of supercapacitors},}\
  }\href@noop {} {\bibfield  {journal} {\bibinfo  {journal} {ChemElectroChem}\
  }\textbf {\bibinfo {volume} {3}},\ \bibinfo {pages} {1429--1436} (\bibinfo
  {year} {2016})}\BibitemShut {NoStop}%
\bibitem [{\citenamefont {Cole}\ and\ \citenamefont
  {Cole}(1941)}]{cole1941dispersion}%
  \BibitemOpen
  \bibfield  {author} {\bibinfo {author} {\bibfnamefont {K.~S.}\ \bibnamefont
  {Cole}}\ and\ \bibinfo {author} {\bibfnamefont {R.~H.}\ \bibnamefont
  {Cole}},\ }\bibfield  {title} {\enquote {\bibinfo {title} {Dispersion and
  absorption in dielectrics i. alternating current characteristics},}\
  }\href@noop {} {\bibfield  {journal} {\bibinfo  {journal} {J. Chem. Phys.}\
  }\textbf {\bibinfo {volume} {9}},\ \bibinfo {pages} {341--351} (\bibinfo
  {year} {1941})}\BibitemShut {NoStop}%
\bibitem [{\citenamefont {Davidson}\ and\ \citenamefont
  {Cole}(1951)}]{davidson1951dielectric}%
  \BibitemOpen
  \bibfield  {author} {\bibinfo {author} {\bibfnamefont {D.~W.}\ \bibnamefont
  {Davidson}}\ and\ \bibinfo {author} {\bibfnamefont {R.~H.}\ \bibnamefont
  {Cole}},\ }\bibfield  {title} {\enquote {\bibinfo {title} {Dielectric
  relaxation in glycerol, propylene glycol, and n-propanol},}\ }\href@noop {}
  {\bibfield  {journal} {\bibinfo  {journal} {J. Chem. Phys.}\ }\textbf
  {\bibinfo {volume} {19}},\ \bibinfo {pages} {1484--1490} (\bibinfo {year}
  {1951})}\BibitemShut {NoStop}%
\bibitem [{\citenamefont {Havriliak}\ and\ \citenamefont
  {Negami}(1966)}]{havriliak1966complex}%
  \BibitemOpen
  \bibfield  {author} {\bibinfo {author} {\bibfnamefont {S.}~\bibnamefont
  {Havriliak}}\ and\ \bibinfo {author} {\bibfnamefont {S.}~\bibnamefont
  {Negami}},\ }\bibfield  {title} {\enquote {\bibinfo {title} {A complex plane
  analysis of $\alpha$-dispersions in some polymer systems},}\ }in\ \href@noop
  {} {\emph {\bibinfo {booktitle} {J. Polym. Sci., Part C: Polym. Symp.}}},\
  Vol.~\bibinfo {volume} {14}\ (\bibinfo {organization} {Wiley Online
  Library},\ \bibinfo {year} {1966})\ pp.\ \bibinfo {pages}
  {99--117}\BibitemShut {NoStop}%
\bibitem [{\citenamefont {Hilfer}(2002{\natexlab{a}})}]{hilfer2002analytical}%
  \BibitemOpen
  \bibfield  {author} {\bibinfo {author} {\bibfnamefont {R.}~\bibnamefont
  {Hilfer}},\ }\bibfield  {title} {\enquote {\bibinfo {title} {Analytical
  representations for relaxation functions of glasses},}\ }\href@noop {}
  {\bibfield  {journal} {\bibinfo  {journal} {J. Non-Cryst. Solids}\ }\textbf
  {\bibinfo {volume} {305}},\ \bibinfo {pages} {122--126} (\bibinfo {year}
  {2002}{\natexlab{a}})}\BibitemShut {NoStop}%
\bibitem [{\citenamefont {Goychuk}(2007)}]{goychuk2007anomalous}%
  \BibitemOpen
  \bibfield  {author} {\bibinfo {author} {\bibfnamefont {I.}~\bibnamefont
  {Goychuk}},\ }\bibfield  {title} {\enquote {\bibinfo {title} {Anomalous
  relaxation and dielectric response},}\ }\href@noop {} {\bibfield  {journal}
  {\bibinfo  {journal} {Phys. Rev. E}\ }\textbf {\bibinfo {volume} {76}},\
  \bibinfo {pages} {040102} (\bibinfo {year} {2007})}\BibitemShut {NoStop}%
\bibitem [{\citenamefont {Allagui}, \citenamefont {Elwakil},\ and\
  \citenamefont {Psychalinos}(2021)}]{cpe}%
  \BibitemOpen
  \bibfield  {author} {\bibinfo {author} {\bibfnamefont {A.}~\bibnamefont
  {Allagui}}, \bibinfo {author} {\bibfnamefont {A.~S.}\ \bibnamefont
  {Elwakil}},\ and\ \bibinfo {author} {\bibfnamefont {C.}~\bibnamefont
  {Psychalinos}},\ }\bibfield  {title} {\enquote {\bibinfo {title} {Decoupling
  the magnitude and phase in a constant phase element},}\ }\href@noop {}
  {\bibfield  {journal} {\bibinfo  {journal} {J. Electroanal. Chem.}\ }\textbf
  {\bibinfo {volume} {888}},\ \bibinfo {pages} {115153} (\bibinfo {year}
  {2021})}\BibitemShut {NoStop}%
\bibitem [{\citenamefont {Garrappa}, \citenamefont {Mainardi},\ and\
  \citenamefont {Maione}(2016)}]{garrappa2016models}%
  \BibitemOpen
  \bibfield  {author} {\bibinfo {author} {\bibfnamefont {R.}~\bibnamefont
  {Garrappa}}, \bibinfo {author} {\bibfnamefont {F.}~\bibnamefont {Mainardi}},\
  and\ \bibinfo {author} {\bibfnamefont {G.}~\bibnamefont {Maione}},\
  }\bibfield  {title} {\enquote {\bibinfo {title} {Models of dielectric
  relaxation based on completely monotone functions},}\ }\href@noop {}
  {\bibfield  {journal} {\bibinfo  {journal} {Fract. Calc. Appl. Anal.}\
  }\textbf {\bibinfo {volume} {19}},\ \bibinfo {pages} {1105--1160} (\bibinfo
  {year} {2016})}\BibitemShut {NoStop}%
\bibitem [{\citenamefont {Prabhakar}(1971)}]{prabhakar1971singular}%
  \BibitemOpen
  \bibfield  {author} {\bibinfo {author} {\bibfnamefont {T.~R.}\ \bibnamefont
  {Prabhakar}},\ }\bibfield  {title} {\enquote {\bibinfo {title} {{A singular
  integral equation with a generalized Mittag Leffler function in the
  kernel}},}\ }\href@noop {} {\bibfield  {journal} {\bibinfo  {journal}
  {Yokohama Math. J.}\ }\textbf {\bibinfo {volume} {19}},\ \bibinfo {pages}
  {7--15} (\bibinfo {year} {1971})}\BibitemShut {NoStop}%
\bibitem [{\citenamefont {Saxena}, \citenamefont {Mathai},\ and\ \citenamefont
  {Haubold}(2004)}]{saxena2004generalized}%
  \BibitemOpen
  \bibfield  {author} {\bibinfo {author} {\bibfnamefont {R.}~\bibnamefont
  {Saxena}}, \bibinfo {author} {\bibfnamefont {A.}~\bibnamefont {Mathai}},\
  and\ \bibinfo {author} {\bibfnamefont {H.}~\bibnamefont {Haubold}},\
  }\bibfield  {title} {\enquote {\bibinfo {title} {On generalized fractional
  kinetic equations},}\ }\href@noop {} {\bibfield  {journal} {\bibinfo
  {journal} {Physica A}\ }\textbf {\bibinfo {volume} {344}},\ \bibinfo {pages}
  {657--664} (\bibinfo {year} {2004})}\BibitemShut {NoStop}%
\bibitem [{\citenamefont {Zeng}\ and\ \citenamefont
  {Chen}(2015)}]{zeng2015global}%
  \BibitemOpen
  \bibfield  {author} {\bibinfo {author} {\bibfnamefont {C.}~\bibnamefont
  {Zeng}}\ and\ \bibinfo {author} {\bibfnamefont {Y.~Q.}\ \bibnamefont
  {Chen}},\ }\bibfield  {title} {\enquote {\bibinfo {title} {{Global Pad\'e
  approximations of the generalized Mittag-Leffler function and its
  inverse}},}\ }\href@noop {} {\bibfield  {journal} {\bibinfo  {journal}
  {Fract. Calc. Appl. Anal.}\ }\textbf {\bibinfo {volume} {18}},\ \bibinfo
  {pages} {1492--1506} (\bibinfo {year} {2015})}\BibitemShut {NoStop}%
\bibitem [{\citenamefont {Hilfer}\ and\ \citenamefont
  {Seybold}(2006)}]{hilfer2006computation}%
  \BibitemOpen
  \bibfield  {author} {\bibinfo {author} {\bibfnamefont {R.}~\bibnamefont
  {Hilfer}}\ and\ \bibinfo {author} {\bibfnamefont {H.}~\bibnamefont
  {Seybold}},\ }\bibfield  {title} {\enquote {\bibinfo {title} {{Computation of
  the generalized Mittag-Leffler function and its inverse in the complex
  plane}},}\ }\href@noop {} {\bibfield  {journal} {\bibinfo  {journal} {Integr.
  Transforms Special Funct.}\ }\textbf {\bibinfo {volume} {17}},\ \bibinfo
  {pages} {637--652} (\bibinfo {year} {2006})}\BibitemShut {NoStop}%
\bibitem [{\citenamefont {Khamzin}, \citenamefont {Nigmatullin},\ and\
  \citenamefont {Popov}(2014)}]{khamzin2014justification}%
  \BibitemOpen
  \bibfield  {author} {\bibinfo {author} {\bibfnamefont {A.}~\bibnamefont
  {Khamzin}}, \bibinfo {author} {\bibfnamefont {R.}~\bibnamefont
  {Nigmatullin}},\ and\ \bibinfo {author} {\bibfnamefont {I.}~\bibnamefont
  {Popov}},\ }\bibfield  {title} {\enquote {\bibinfo {title} {Justification of
  the empirical laws of the anomalous dielectric relaxation in the framework of
  the memory function formalism},}\ }\href@noop {} {\bibfield  {journal}
  {\bibinfo  {journal} {Fract. Calc. Appl. Anal.}\ }\textbf {\bibinfo {volume}
  {17}},\ \bibinfo {pages} {247--258} (\bibinfo {year} {2014})}\BibitemShut
  {NoStop}%
\bibitem [{\citenamefont {Podlubny}(1998)}]{podlubny1998fractional}%
  \BibitemOpen
  \bibfield  {author} {\bibinfo {author} {\bibfnamefont {I.}~\bibnamefont
  {Podlubny}},\ }\href@noop {} {\emph {\bibinfo {title} {Fractional
  differential equations: an introduction to fractional derivatives, fractional
  differential equations, to methods of their solution and some of their
  applications}}}\ (\bibinfo  {publisher} {Elsevier},\ \bibinfo {year}
  {1998})\BibitemShut {NoStop}%
\bibitem [{\citenamefont {Nigmatullin}(1984)}]{nigmatullin1984theory}%
  \BibitemOpen
  \bibfield  {author} {\bibinfo {author} {\bibfnamefont {R.}~\bibnamefont
  {Nigmatullin}},\ }\bibfield  {title} {\enquote {\bibinfo {title} {On the
  theory of relaxation for systems with "remnant" memory},}\ }\href@noop {}
  {\bibfield  {journal} {\bibinfo  {journal} {Phys. Status Solidi B}\ }\textbf
  {\bibinfo {volume} {124}},\ \bibinfo {pages} {389--393} (\bibinfo {year}
  {1984})}\BibitemShut {NoStop}%
\bibitem [{\citenamefont {Allagui}, \citenamefont {Zhang},\ and\ \citenamefont
  {Elwakil}(2018)}]{memoryAPL}%
  \BibitemOpen
  \bibfield  {author} {\bibinfo {author} {\bibfnamefont {A.}~\bibnamefont
  {Allagui}}, \bibinfo {author} {\bibfnamefont {D.}~\bibnamefont {Zhang}},\
  and\ \bibinfo {author} {\bibfnamefont {A.~S.}\ \bibnamefont {Elwakil}},\
  }\bibfield  {title} {\enquote {\bibinfo {title} {Short-term memory in
  electric double-layer capacitors},}\ }\href@noop {} {\bibfield  {journal}
  {\bibinfo  {journal} {Appl. Phys. Lett.}\ }\textbf {\bibinfo {volume}
  {113}},\ \bibinfo {pages} {253901--5} (\bibinfo {year} {2018})}\BibitemShut
  {NoStop}%
\bibitem [{\citenamefont {Allagui}\ \emph {et~al.}(2020)\citenamefont
  {Allagui}, \citenamefont {Zhang}, \citenamefont {Khakpour}, \citenamefont
  {Elwakil},\ and\ \citenamefont {Wang}}]{memQ}%
  \BibitemOpen
  \bibfield  {author} {\bibinfo {author} {\bibfnamefont {A.}~\bibnamefont
  {Allagui}}, \bibinfo {author} {\bibfnamefont {D.}~\bibnamefont {Zhang}},
  \bibinfo {author} {\bibfnamefont {I.}~\bibnamefont {Khakpour}}, \bibinfo
  {author} {\bibfnamefont {A.~S.}\ \bibnamefont {Elwakil}},\ and\ \bibinfo
  {author} {\bibfnamefont {C.}~\bibnamefont {Wang}},\ }\bibfield  {title}
  {\enquote {\bibinfo {title} {Quantification of memory in fractional-order
  capacitors},}\ }\href@noop {} {\bibfield  {journal} {\bibinfo  {journal} {J.
  Phys. D}\ }\textbf {\bibinfo {volume} {53}} (\bibinfo {year}
  {2020})}\BibitemShut {NoStop}%
\bibitem [{\citenamefont {Allagui}\ and\ \citenamefont
  {Elwakil}(2021)}]{allagui2021possibility}%
  \BibitemOpen
  \bibfield  {author} {\bibinfo {author} {\bibfnamefont {A.}~\bibnamefont
  {Allagui}}\ and\ \bibinfo {author} {\bibfnamefont {A.~S.}\ \bibnamefont
  {Elwakil}},\ }\bibfield  {title} {\enquote {\bibinfo {title} {Possibility of
  information encoding/decoding using the memory effect in fractional-order
  capacitive devices},}\ }\href@noop {} {\bibfield  {journal} {\bibinfo
  {journal} {Sci. Rep.}\ }\textbf {\bibinfo {volume} {11}},\ \bibinfo {pages}
  {1--7} (\bibinfo {year} {2021})}\BibitemShut {NoStop}%
\bibitem [{\citenamefont {Rosa}\ and\ \citenamefont
  {de~Oliveira}(2015)}]{rosa2015relaxation}%
  \BibitemOpen
  \bibfield  {author} {\bibinfo {author} {\bibfnamefont {C.}~\bibnamefont
  {Rosa}}\ and\ \bibinfo {author} {\bibfnamefont {E.~C.}\ \bibnamefont
  {de~Oliveira}},\ }\bibfield  {title} {\enquote {\bibinfo {title} {Relaxation
  equations: fractional models},}\ }\href@noop {} {\bibfield  {journal}
  {\bibinfo  {journal} {Journal of Physical Mathematics}\ }\textbf {\bibinfo
  {volume} {6}},\ \bibinfo {pages} {1--7} (\bibinfo {year} {2015})}\BibitemShut
  {NoStop}%
\bibitem [{\citenamefont {Kilbas}, \citenamefont {Saigo},\ and\ \citenamefont
  {Saxena}(2004)}]{kilbas2004generalized}%
  \BibitemOpen
  \bibfield  {author} {\bibinfo {author} {\bibfnamefont {A.~A.}\ \bibnamefont
  {Kilbas}}, \bibinfo {author} {\bibfnamefont {M.}~\bibnamefont {Saigo}},\ and\
  \bibinfo {author} {\bibfnamefont {R.~K.}\ \bibnamefont {Saxena}},\ }\bibfield
   {title} {\enquote {\bibinfo {title} {{Generalized Mittag-Leffler function
  and generalized fractional calculus operators}},}\ }\href@noop {} {\bibfield
  {journal} {\bibinfo  {journal} {Integr. Transforms Special Funct.}\ }\textbf
  {\bibinfo {volume} {15}},\ \bibinfo {pages} {31--49} (\bibinfo {year}
  {2004})}\BibitemShut {NoStop}%
\bibitem [{\citenamefont {Weron}, \citenamefont {Jurlewicz},\ and\
  \citenamefont {Magdziarz}(2005)}]{weron2005havriliak}%
  \BibitemOpen
  \bibfield  {author} {\bibinfo {author} {\bibfnamefont {K.}~\bibnamefont
  {Weron}}, \bibinfo {author} {\bibfnamefont {A.}~\bibnamefont {Jurlewicz}},\
  and\ \bibinfo {author} {\bibfnamefont {M.}~\bibnamefont {Magdziarz}},\
  }\bibfield  {title} {\enquote {\bibinfo {title} {Havriliak--negami response
  in the framework of the continuous-time random walk},}\ }\href@noop {}
  {\bibfield  {journal} {\bibinfo  {journal} {Acta Phys. Pol. B}\ }\textbf
  {\bibinfo {volume} {36}},\ \bibinfo {pages} {1855--1868} (\bibinfo {year}
  {2005})}\BibitemShut {NoStop}%
\bibitem [{\citenamefont {Hilfer}(2002{\natexlab{b}})}]{hilfer2002h}%
  \BibitemOpen
  \bibfield  {author} {\bibinfo {author} {\bibfnamefont {R.}~\bibnamefont
  {Hilfer}},\ }\bibfield  {title} {\enquote {\bibinfo {title} {H-function
  representations for stretched exponential relaxation and non-debye
  susceptibilities in glassy systems},}\ }\href@noop {} {\bibfield  {journal}
  {\bibinfo  {journal} {Phys. Rev. E}\ }\textbf {\bibinfo {volume} {65}},\
  \bibinfo {pages} {061510} (\bibinfo {year} {2002}{\natexlab{b}})}\BibitemShut
  {NoStop}%
\bibitem [{\citenamefont {Borges}(1998)}]{borges1998q}%
  \BibitemOpen
  \bibfield  {author} {\bibinfo {author} {\bibfnamefont {E.~P.}\ \bibnamefont
  {Borges}},\ }\bibfield  {title} {\enquote {\bibinfo {title} {On a
  q-generalization of circular and hyperbolic functions},}\ }\href@noop {}
  {\bibfield  {journal} {\bibinfo  {journal} {J. Phys. A: Math. Gen.}\ }\textbf
  {\bibinfo {volume} {31}},\ \bibinfo {pages} {5281} (\bibinfo {year}
  {1998})}\BibitemShut {NoStop}%
\bibitem [{\citenamefont {Tsallis}(1988)}]{tsallis1988possible}%
  \BibitemOpen
  \bibfield  {author} {\bibinfo {author} {\bibfnamefont {C.}~\bibnamefont
  {Tsallis}},\ }\bibfield  {title} {\enquote {\bibinfo {title} {{Possible
  generalization of Boltzmann-Gibbs statistics}},}\ }\href@noop {} {\bibfield
  {journal} {\bibinfo  {journal} {J. Stat. Phys.}\ }\textbf {\bibinfo {volume}
  {52}},\ \bibinfo {pages} {479--487} (\bibinfo {year} {1988})}\BibitemShut
  {NoStop}%
\bibitem [{\citenamefont {Tsallis}(2004)}]{tsallis2004should}%
  \BibitemOpen
  \bibfield  {author} {\bibinfo {author} {\bibfnamefont {C.}~\bibnamefont
  {Tsallis}},\ }\bibfield  {title} {\enquote {\bibinfo {title} {What should a
  statistical mechanics satisfy to reflect nature?}}\ }\href@noop {} {\bibfield
   {journal} {\bibinfo  {journal} {Physica D}\ }\textbf {\bibinfo {volume}
  {193}},\ \bibinfo {pages} {3--34} (\bibinfo {year} {2004})}\BibitemShut
  {NoStop}%
\bibitem [{\citenamefont {Beck}\ and\ \citenamefont
  {Cohen}(2003)}]{beck2003superstatistics}%
  \BibitemOpen
  \bibfield  {author} {\bibinfo {author} {\bibfnamefont {C.}~\bibnamefont
  {Beck}}\ and\ \bibinfo {author} {\bibfnamefont {E.}~\bibnamefont {Cohen}},\
  }\bibfield  {title} {\enquote {\bibinfo {title} {Superstatistics},}\
  }\href@noop {} {\bibfield  {journal} {\bibinfo  {journal} {Physica A}\
  }\textbf {\bibinfo {volume} {322}},\ \bibinfo {pages} {267--275} (\bibinfo
  {year} {2003})}\BibitemShut {NoStop}%
\bibitem [{\citenamefont {Allagui}, \citenamefont {Benaoum},\ and\
  \citenamefont {Wang}(2022)}]{Allagui:2022ue}%
  \BibitemOpen
  \bibfield  {author} {\bibinfo {author} {\bibfnamefont {A.}~\bibnamefont
  {Allagui}}, \bibinfo {author} {\bibfnamefont {H.}~\bibnamefont {Benaoum}},\
  and\ \bibinfo {author} {\bibfnamefont {C.}~\bibnamefont {Wang}},\ }\bibfield
  {title} {\enquote {\bibinfo {title} {{Deformed Butler--Volmer Models for
  Convex Semilogarithmic Current-Overpotential Profiles of Li-ion
  Batteries}},}\ }\href@noop {} {\bibfield  {journal} {\bibinfo  {journal} {J.
  Phys. Chem. C}\ }\textbf {\bibinfo {volume} {126}},\ \bibinfo {pages}
  {3029--3036} (\bibinfo {year} {2022})}\BibitemShut {NoStop}%
\bibitem [{\citenamefont {Allagui}, \citenamefont {Benaoum},\ and\
  \citenamefont {Olendski}(2021)}]{allagui2021gouy}%
  \BibitemOpen
  \bibfield  {author} {\bibinfo {author} {\bibfnamefont {A.}~\bibnamefont
  {Allagui}}, \bibinfo {author} {\bibfnamefont {H.}~\bibnamefont {Benaoum}},\
  and\ \bibinfo {author} {\bibfnamefont {O.}~\bibnamefont {Olendski}},\
  }\bibfield  {title} {\enquote {\bibinfo {title} {{On the Gouy-Chapman-Stern
  model of the electrical double-layer structure with a generalized Boltzmann
  factor}},}\ }\href@noop {} {\bibfield  {journal} {\bibinfo  {journal}
  {Physica A}\ ,\ \bibinfo {pages} {126252}} (\bibinfo {year}
  {2021})}\BibitemShut {NoStop}%
\bibitem [{\citenamefont {Borges}(2004)}]{borges2004possible}%
  \BibitemOpen
  \bibfield  {author} {\bibinfo {author} {\bibfnamefont {E.~P.}\ \bibnamefont
  {Borges}},\ }\bibfield  {title} {\enquote {\bibinfo {title} {A possible
  deformed algebra and calculus inspired in nonextensive thermostatistics},}\
  }\href@noop {} {\bibfield  {journal} {\bibinfo  {journal} {Physica A}\
  }\textbf {\bibinfo {volume} {340}},\ \bibinfo {pages} {95--101} (\bibinfo
  {year} {2004})}\BibitemShut {NoStop}%
\bibitem [{\citenamefont {Mathai}\ and\ \citenamefont
  {Moschopoulos}(2012)}]{mathai2012pathway}%
  \BibitemOpen
  \bibfield  {author} {\bibinfo {author} {\bibfnamefont {A.}~\bibnamefont
  {Mathai}}\ and\ \bibinfo {author} {\bibfnamefont {P.}~\bibnamefont
  {Moschopoulos}},\ }\bibfield  {title} {\enquote {\bibinfo {title} {A pathway
  idea for model building},}\ }\href@noop {} {\bibfield  {journal} {\bibinfo
  {journal} {Journal of statistics applications \& probability}\ }\textbf
  {\bibinfo {volume} {1}},\ \bibinfo {pages} {15} (\bibinfo {year}
  {2012})}\BibitemShut {NoStop}%
\bibitem [{\citenamefont {Mathai}(2005)}]{mathai2005pathway}%
  \BibitemOpen
  \bibfield  {author} {\bibinfo {author} {\bibfnamefont {A.}~\bibnamefont
  {Mathai}},\ }\bibfield  {title} {\enquote {\bibinfo {title} {A pathway to
  matrix-variate gamma and normal densities},}\ }\href@noop {} {\bibfield
  {journal} {\bibinfo  {journal} {Linear Algebra Appl.}\ }\textbf {\bibinfo
  {volume} {396}},\ \bibinfo {pages} {317--328} (\bibinfo {year}
  {2005})}\BibitemShut {NoStop}%
\bibitem [{\citenamefont {Mathai}\ and\ \citenamefont
  {Haubold}(2007)}]{mathai2007pathway}%
  \BibitemOpen
  \bibfield  {author} {\bibinfo {author} {\bibfnamefont {A.}~\bibnamefont
  {Mathai}}\ and\ \bibinfo {author} {\bibfnamefont {H.~J.}\ \bibnamefont
  {Haubold}},\ }\bibfield  {title} {\enquote {\bibinfo {title} {Pathway model,
  superstatistics, tsallis statistics, and a generalized measure of entropy},}\
  }\href@noop {} {\bibfield  {journal} {\bibinfo  {journal} {Physica A}\
  }\textbf {\bibinfo {volume} {375}},\ \bibinfo {pages} {110--122} (\bibinfo
  {year} {2007})}\BibitemShut {NoStop}%
\bibitem [{\citenamefont {Mathai}\ and\ \citenamefont
  {Haubold}(2015)}]{mathai2015stochastic}%
  \BibitemOpen
  \bibfield  {author} {\bibinfo {author} {\bibfnamefont {A.~M.}\ \bibnamefont
  {Mathai}}\ and\ \bibinfo {author} {\bibfnamefont {H.~J.}\ \bibnamefont
  {Haubold}},\ }\bibfield  {title} {\enquote {\bibinfo {title} {Stochastic
  processes via the pathway model},}\ }\href@noop {} {\bibfield  {journal}
  {\bibinfo  {journal} {Entropy}\ }\textbf {\bibinfo {volume} {17}},\ \bibinfo
  {pages} {2642--2654} (\bibinfo {year} {2015})}\BibitemShut {NoStop}%
\bibitem [{\citenamefont {Sebastian}, \citenamefont {S~Nair},\ and\
  \citenamefont {P~Joseph}(2015)}]{sebastian2015overview}%
  \BibitemOpen
  \bibfield  {author} {\bibinfo {author} {\bibfnamefont {N.}~\bibnamefont
  {Sebastian}}, \bibinfo {author} {\bibfnamefont {S.}~\bibnamefont {S~Nair}},\
  and\ \bibinfo {author} {\bibfnamefont {D.}~\bibnamefont {P~Joseph}},\
  }\bibfield  {title} {\enquote {\bibinfo {title} {An overview of the pathway
  idea and its applications in statistical and physical sciences},}\
  }\href@noop {} {\bibfield  {journal} {\bibinfo  {journal} {Axioms}\ }\textbf
  {\bibinfo {volume} {4}},\ \bibinfo {pages} {530--553} (\bibinfo {year}
  {2015})}\BibitemShut {NoStop}%
\bibitem [{\citenamefont {Jagannathan}\ and\ \citenamefont
  {Khan}(2020)}]{jagannathan2020deformed}%
  \BibitemOpen
  \bibfield  {author} {\bibinfo {author} {\bibfnamefont {R.}~\bibnamefont
  {Jagannathan}}\ and\ \bibinfo {author} {\bibfnamefont {S.~A.}\ \bibnamefont
  {Khan}},\ }\bibfield  {title} {\enquote {\bibinfo {title} {{On the Deformed
  Oscillator and the Deformed Derivative Associated with the Tsallis
  q-exponential}},}\ }\href@noop {} {\bibfield  {journal} {\bibinfo  {journal}
  {Int. J. Theor. Phys.}\ }\textbf {\bibinfo {volume} {59}},\ \bibinfo {pages}
  {2647--2669} (\bibinfo {year} {2020})}\BibitemShut {NoStop}%
\bibitem [{\citenamefont {Jordehi}(2016)}]{jordehi2016time}%
  \BibitemOpen
  \bibfield  {author} {\bibinfo {author} {\bibfnamefont {A.~R.}\ \bibnamefont
  {Jordehi}},\ }\bibfield  {title} {\enquote {\bibinfo {title} {Time varying
  acceleration coefficients particle swarm optimisation (tvacpso): A new
  optimisation algorithm for estimating parameters of pv cells and modules},}\
  }\href@noop {} {\bibfield  {journal} {\bibinfo  {journal} {Energy Convers.
  Manage.}\ }\textbf {\bibinfo {volume} {129}},\ \bibinfo {pages} {262--274}
  (\bibinfo {year} {2016})}\BibitemShut {NoStop}%
\bibitem [{\citenamefont {Fouda}\ \emph {et~al.}(2020)\citenamefont {Fouda},
  \citenamefont {Allagui}, \citenamefont {Elwakil}, \citenamefont {Das},
  \citenamefont {Psychalinos},\ and\ \citenamefont {Radwan}}]{QV}%
  \BibitemOpen
  \bibfield  {author} {\bibinfo {author} {\bibfnamefont {M.~E.}\ \bibnamefont
  {Fouda}}, \bibinfo {author} {\bibfnamefont {A.}~\bibnamefont {Allagui}},
  \bibinfo {author} {\bibfnamefont {A.~S.}\ \bibnamefont {Elwakil}}, \bibinfo
  {author} {\bibfnamefont {S.}~\bibnamefont {Das}}, \bibinfo {author}
  {\bibfnamefont {C.}~\bibnamefont {Psychalinos}},\ and\ \bibinfo {author}
  {\bibfnamefont {A.~G.}\ \bibnamefont {Radwan}},\ }\bibfield  {title}
  {\enquote {\bibinfo {title} {Nonlinear charge-voltage relationship in
  constant phase element},}\ }\href@noop {} {\bibfield  {journal} {\bibinfo
  {journal} {AEU Int. J. Electron. Commun.}\ }\textbf {\bibinfo {volume} {117}}
  (\bibinfo {year} {2020})}\BibitemShut {NoStop}%
\bibitem [{\citenamefont {Mathai}, \citenamefont {Saxena},\ and\ \citenamefont
  {Haubold}(2009)}]{mathai2009h}%
  \BibitemOpen
  \bibfield  {author} {\bibinfo {author} {\bibfnamefont {A.~M.}\ \bibnamefont
  {Mathai}}, \bibinfo {author} {\bibfnamefont {R.~K.}\ \bibnamefont {Saxena}},\
  and\ \bibinfo {author} {\bibfnamefont {H.~J.}\ \bibnamefont {Haubold}},\
  }\href@noop {} {\emph {\bibinfo {title} {The H-function: theory and
  applications}}}\ (\bibinfo  {publisher} {Springer Science \& Business
  Media},\ \bibinfo {year} {2009})\BibitemShut {NoStop}%
\bibitem [{\citenamefont {Mathai}\ \emph {et~al.}(1978)\citenamefont {Mathai},
  \citenamefont {Saxena}, \citenamefont {Saxena} \emph {et~al.}}]{mathai1978h}%
  \BibitemOpen
  \bibfield  {author} {\bibinfo {author} {\bibfnamefont {A.~M.}\ \bibnamefont
  {Mathai}}, \bibinfo {author} {\bibfnamefont {R.~K.}\ \bibnamefont {Saxena}},
  \bibinfo {author} {\bibfnamefont {R.~K.}\ \bibnamefont {Saxena}}, \emph
  {et~al.},\ }\href@noop {} {\emph {\bibinfo {title} {The H-function with
  applications in statistics and other disciplines}}}\ (\bibinfo  {publisher}
  {John Wiley \& Sons},\ \bibinfo {year} {1978})\BibitemShut {NoStop}%
\end{thebibliography}
 
%

 \end{document}